\documentclass[fleqn,usenatbib]{mnras}

\usepackage{newtxtext,newtxmath}
\usepackage[T1]{fontenc}

\DeclareRobustCommand{\VAN}[3]{#2}
\let\VANthebibliography\thebibliography
\def\thebibliography{\DeclareRobustCommand{\VAN}[3]{##3}\VANthebibliography}

\usepackage{graphicx}	

\usepackage{amsmath}	
\usepackage{amssymb}	
\usepackage{dsfont}
\usepackage{algorithm}
\usepackage{algpseudocode}
\usepackage[mathscr]{euscript}
\usepackage{microtype}
\usepackage{subfigure}
\usepackage{booktabs}
\usepackage{hyperref}
\usepackage{lastpage}

\title[Generative Models for High-Dimensional Field-Level Inference]{Learning the Universe: Posterior Reliability of Neural Generative Models in High-Dimensional Field-Level Inference of Cosmic Initial Conditions}

\author[L. Doeser, J. Jasche]{
Ludvig Doeser,$^{1,2}$\thanks{E-mail: ludvig.doeser@fysik.su.se}
Jens Jasche$^{1}$
\\
$^{1}$The Oskar Klein Centre, Department of Physics, Stockholm University, AlbaNova University Centre, SE 106 91 Stockholm, Sweden \\
$^{2}$Center for Computational Astrophysics, Flatiron Institute, 162 5th Avenue, New York, NY 10010, USA
}

\date{Accepted XXX. Received YYY; in original form ZZZ}

\pubyear{2026}

\begin{document}
\label{firstpage}
\pagerange{\pageref{firstpage}--\pageref{lastpage}}
\maketitle

\begin{abstract}
Accurate posterior estimation is central to scientific inference, as uncertainties determine what can be reliably learned from observational data. While Markov chain Monte Carlo methods provide asymptotic convergence guarantees, they are computationally demanding in high-dimensional settings. Neural network–based generative models for entire discretized $3$D fields enable fast amortized inference but often lack convergence guarantees and principled accuracy assessment. Using Hamiltonian Monte Carlo to obtain reference posterior samples, we conduct a controlled field-level evaluation of an implicit generative model (Stochastic Interpolants) and an explicit likelihood-based model (GLOW normalizing flows). This comparison, unavailable in typical applications, enables the detection of posterior geometry failures that standard metrics cannot capture. As a case study, we consider the cosmological inverse problem of inferring cosmic initial conditions from present-day large-scale structure. To match the precision of modern cosmological data, this problem increasingly relies on complex, non-linear, and non-differentiable simulators, which are incompatible with gradient-based inference frameworks. Generative models offer a route to address these challenges, provided their inferred posteriors are reliable. In this work, we show that matching posterior means, marginal distributions, or achieving high cross-correlation does not imply correct uncertainty structure, as revealed by posterior variance fields and sample-based evaluations. Through this work, we aim to raise awareness of the challenges of uncertainty estimation in high-dimensional field-level settings, highlighting the importance of careful design and validation of neural generative approaches for scientific applications.
\end{abstract}


\section{Introduction}
Modern cosmological data, from e.g. DESI \citep{DESICollaboration2016}, Euclid \citep{Laureijs2011,Amendola2018}, LSST \citep{LSSTScienceCollaboration2009,LSSTDarkEnergyScienceCollaboration2012,Ivezic2019}, SPHEREx \citep{doreCosmologySPHEREXAllSky2014,doreScienceImpactsSPHEREx2018}, and the Subaru Prime Focus Spectrograph \citep{Takada2012}, increasingly require high-fidelity modeling to accurately describe non-linear structure formation. This can be achieved by incorporating numerical simulations to fully exploit the information content of the data within field-level inference of the initial conditions of the Universe \citep[e.g.][]{Jasche2012,wangRECONSTRUCTINGINITIALDENSITY2013,Kitaura2013,Jasche2019,Ata2020,McAlpine2025}. Improving physical realism, however, also introduces substantial complexity: high-fidelity simulators are often non-linear, computationally expensive, and may include non-differentiable components. 

Neural networks offer a flexible class of models that can be leveraged in this context due to their strong expressive power \citep{funahashiApproximateRealizationContinuous1989,cybenkoApproximationSuperpositionsSigmoidal1989}
and ability to approximate continuous functions to arbitrary accuracy \citep{hornikMultilayerFeedforwardNetworks1989,luExpressivePowerNeural2017,augustineSurveyUniversalApproximation2024}. Combined with their computational efficiency once trained, this has led to the rapid adoption of neural network–based methods in scientific inference. Within inference of cosmic initial conditions, they can accelerate simulations \citep{Modi2018,He2019,kaushalNECOLAUniversalFieldlevel2021,doeser2024,Jamieson2022,Jamieson2024,Bartlett2024}, directly improve the reconstruction \citep{Jindal2023,Chen2023,Chen2024,Bottema2025,flossImprovingConstraintsPrimordial2024,Parker2025, bayerFieldLevelInferenceGalaxies2026} or be used as neural optimizers \citep{Modi2021a,Doeser2025}.

An alternative approach that leverages both forward simulators and neural networks to perform inference and obtain posterior distributions is simulation-based inference (SBI) \citep{papamakariosFast$e$freeInference2018,cranmerFrontierSimulationbasedInference2020}. Instead of requiring an analytic data likelihood, SBI methods rely on forward simulations: by sampling input parameters from the prior and running the simulator, one obtains joint samples from the parameter–data distribution. Modern neural generative models for SBI, including normalizing flows \citep{rezendeVariationalInferenceNormalizing2016,papamakariosNormalizingFlowsProbabilistic2021}, diffusion and score-based models \citep{hoDenoisingDiffusionProbabilistic2020,songScoreBasedGenerativeModeling2021}, and flow-matching approaches \citep{lipmanFlowMatchingGenerative2023}, can then be trained to infer the unknown parameters or full field by approximating either the likelihood or the posterior \citep[e.g.][]{alsing2019,brehmerMiningGoldImplicit2020,goncalvesTrainingDeepNeural2020,daxFlowMatchingScalable2023,hoLtUILIAllinOneFramework2024}.

In natural science applications, however, the goals of inference differ from those in typical generative modeling settings \citep[as highlighted in e.g.][]{hermansLikelihoodfreeMCMCAmortized2020,carzonTrustworthyScientificInference2026}. Whereas many machine-learning tasks prioritize perceptual fidelity or sample realism, scientific inverse problems require \emph{accurate uncertainty quantification}. The posterior mean describes the best estimate of the underlying parameter(s)/field(s), but the variance, covariance, and higher-order moments determine the range of plausible solutions. These play a central role in correct data interpretation and downstream decision-making. In these settings, generative models must not only produce plausible reconstructions, but also accurately describe the \emph{full posterior geometry}. A standard approach for validating inferred distributions in SBI is the use of coverage-based diagnostics \citep[e.g.][]{taltsValidatingBayesianInference2020,lemosSamplingBasedAccuracyTesting2023}. Crucially, however, such diagnostics probe calibration only in an averaged sense, and it has recently been shown by \citet{alokdaCoverageNotEnough2026} that satisfying these criteria does not necessarily guarantee reliable posterior structure for a fixed realization, which is ultimately the regime relevant for scientific applications.

One important class of SBI methods is \emph{field-level} SBI, in which entire three-dimensional ($3$D) discretized fields are inferred, with each voxel corresponding to a parameter, typically leading to total dimensionalities of $10^4$–$10^9$. This high-dimensional setting introduces a range of challenges, such as generating sufficiently diverse training data, learning efficiently from a limited number of simulations, and validating posterior estimates. In cosmology, this approach has been applied to the inference of cosmic initial conditions \citep[e.g.][]{List2023,Legin2023,cuesta2024joint,Savchenko2024,Savchenko2025,bayerFieldLevelInferenceGalaxies2026}. In this work, we consider inference of cosmic initial conditions using an efficient, differentiable simulator. This enables both the generation of large training datasets and reference posterior samples via Hamiltonian Monte Carlo (HMC), which provides asymptotic convergence guarantees. While field-level SBI is ultimately aimed at settings where the target posterior cannot be sampled with gradient-based methods such as HMC (e.g., due to non-differentiable simulator components), we here consider a controlled scenario in which it can. This allows us to directly compare explicit and implicit neural generative models against an HMC reference posterior and assess their ability to capture the detailed posterior geometry conditioned on a single realization.

In this work, we consider two classes of generative models that are increasingly used for high-dimensional posterior learning in astrophysics and cosmology: GLOW normalizing flows \citep[as used in e.g.][]{friedmanHIGlowConditionalNormalizing2022,rouhiainenNormalizingFlowsRandom2021b,daiMultiscaleFlowRobust2024} and Stochastic Interpolants (SI) \citep[e.g. in][]{cuesta2024joint,sabtiGenerativeModelingApproach2024,riverosConditionalDiffusionFlowModels2025,horowitzBaryonBridgeStochasticInterpolant2025}. In particular, we employ a conditional GLOW model \citep{kingmaGlowGenerativeFlow2018,luStructuredOutputLearning2020a}, a well-established and scalable architecture for $3$D volumetric data. For implicit generative modeling, we adopt the SI framework \citep{Albergo2025}, which enables training via flow-matching objectives that learn a transport field between distributions. By introducing noise into the interpolant, this approach is closely connected to diffusion-based models \citep{hoDenoisingDiffusionProbabilistic2020,songScoreBasedGenerativeModeling2021}, while retaining flexibility in the choice of interpolation path and base distribution.

This work is carried out within the broader scientific programme of the Simons Collaboration on Learning the Universe (LtU)\footnote{\hyperlink{https://learning-the-universe.org/}{https://learning-the-universe.org/}}, which seeks to learn the cosmological parameters and initial conditions of the Universe by jointly leveraging machine learning, Bayesian forward modelling, and state-of-the-art simulations. A central component of this effort is high-resolution field-level inference, yielding physically consistent posterior ensembles of initial conditions and corresponding present-day structures with proper uncertainty quantification \citep[e.g.][released along this work]{ManticoreDeep2026}. Rather than targeting inference from observational data, this study provides a controlled comparison of neural posterior methods with an established Bayesian forward modelling approach. By assessing the consistency of the resulting posterior structure, we aim to evaluate their reliability for application in high-dimensional field-level inference with real data.

The paper is structured as follows. Section~\ref{sec:theory} introduces simulation-based inference and the challenges associated with high-dimensional problems. Section~\ref{sec:cosmo_case} describes the cosmological setting, followed by section~\ref{sec:gen_models}, which presents the generative models used for inference. In section~\ref{sec:results}, we present a quantitative comparison of the neural posteriors against the reference HMC posterior. Finally, section~\ref{sec:discussion} and section~\ref{sec:conclusion} discusses and summarizes our findings.

\section{Simulation-Based Inference: Formulation and Key Challenges at Field Level}
\label{sec:theory}

We consider the standard setting of \emph{simulation-based inference} (SBI), where training data consist of joint samples
\begin{equation}
    (\boldsymbol{x}, \boldsymbol{y}) \sim \pi(\boldsymbol{x})\, \pi(\boldsymbol{y} \mid \boldsymbol{x}),
    \label{eq:SBI}
\end{equation}
with $\pi(\boldsymbol{x})$ a known prior and $\pi(\boldsymbol{y}\mid \boldsymbol{x})$ a likelihood accessible through simulation. In the field-level setting, both the input $\boldsymbol{x}$ and the simulated data $\boldsymbol{y}$ are $3$D gridded fields. The goal is to train a generative model $q_{\theta}$ with parameters $\theta$ to approximate the posterior distribution $\pi(\boldsymbol{x}\mid \boldsymbol{y}_o)$ for a fixed observation $\boldsymbol{y}_o$ to answer: \textit{what is the set of plausible initial conditions that could have given rise to the data?}

\subsection{Complexities for high-dimensional SBI}

Learning a posterior distribution with a generative model in a high-dimensional setting from finite simulation data constitutes an instance of empirical risk minimization, where the neural model parameters are optimized to minimize the average loss evaluated on the available data \citep[see e.g.][]{Goodfellow2016}. The resulting generalization error depends on several coupled aspects: (i) training dataset size $N$, (ii) representation capacity of the generative model, and (iii) optimization effectiveness and the training objective.

In this work, we consider both explicit and implicit neural generative models for $3$D fields to assess how architectural choices, training data variants, and training objectives affect posterior approximation. The generative models and their associated training objectives are introduced in section~\ref{sec:gen_models}. In the following, we discuss challenges related to sample complexity. 

\begin{table*}
\centering
\caption{
Training data regimes and corresponding model training formulations considered in this work.
In the \textit{rnd} setting, models are trained on samples from the joint distribution.
In the \textit{hmc} setting, models are trained directly on posterior samples at a fixed observation $\boldsymbol{y}_o$, corresponding to distributional fitting. $^*$To investigate training data convergence, we train one model with $262\,144$ samples (see section~\ref{sec:training_data_convergence}).
}
\label{tab:training_regimes}
\begin{tabular}{lllll}
\hline
Setting & Training data & $N_{\mathrm{samples}}$ & GLOW training & SI training \\
\hline
\textit{rnd }
& $(\boldsymbol{x},\boldsymbol{y}) \sim \pi(\boldsymbol{x})\,\pi(\boldsymbol{y}\mid\boldsymbol{x})$
& $59\,880^*$ & Conditional density estimation $q_{\mathrm{GLOW}_\mathrm{rnd}}(\boldsymbol{x}\mid\boldsymbol{y})$
& Interpolating from $\pi(\boldsymbol{y}\mid\boldsymbol{x})$ to $q_{\mathrm{SI}_\mathrm{rnd}}(\boldsymbol{x})$ \\

\textit{hmc }
& $\boldsymbol{x} \sim \pi_{\mathrm{HMC}}(\boldsymbol{x}\mid\boldsymbol{y}_o)$
& $59\,880$ &  Unconditional density est. $q_{\mathrm{GLOW}_\mathrm{hmc}}(\boldsymbol{x}\mid\boldsymbol{y}_o)$
& Interpolating from $\mathcal{N}(\boldsymbol{y}_o,\tfrac{1}{10}\mathbf{I})$ to $q_{\mathrm{SI}_\mathrm{hmc}}(\boldsymbol{x}\mid\boldsymbol{y}_o)$ \\
\hline
\end{tabular}
\end{table*}

\subsection{Sparse coverage near a fixed observation}
Training data in SBI are typically drawn from the joint distribution $\pi(\boldsymbol{x},\boldsymbol{y})$, while inference targets the posterior at a fixed observation $\boldsymbol{y}_o$. The coverage of the parameter region of interest is set by the sample size and the dimensionality $d=\dim(\boldsymbol{x})$. In our setting, $\boldsymbol{x}\sim\mathcal{N}(\mathbf{0},\mathbf{I}_d)$ (see section~\ref{sec:cosmo_case}) and as $d$ increases samples concentrate on a thin shell of radius $\sqrt{d}$ for any fixed $\varepsilon > 0$. Yet, any two independent samples remain separated by distances 
$\|\boldsymbol{x} - \boldsymbol{x}'\|_2^2 \sim 2d$, 
while their inner products satisfy 
$\boldsymbol{x}^\top \boldsymbol{x}' \sim 0$, 
implying that random vectors become nearly orthogonal in high dimensions \citep[see e.g.][]{vershyninHighDimensionalProbabilityIntroduction2018a}. 

As a result, finite training sets typically provide sparse coverage in high-dimensional parameter space. This, in turn, leads to low coverage of the data distribution; conditioning on a fixed observation $\boldsymbol{y}_o$ makes draws with $\boldsymbol{y}\approx\boldsymbol{y}_o$ increasingly unlikely as $d$ grows. 

Field-level posterior learning in high dimensions is thus fundamentally challenging. Even with large simulation budgets, training data are sparse relative to the typical set of the posterior. Quantifying the simulation cost required for reliable SBI is an active area of research \citep[e.g.][]{bairagiHowManySimulations2025}, and several approaches have been proposed to improve sample efficiency, including sequential, adaptive, and multi-fidelity strategies that focus simulations on informative regions of the parameter space \citep{papamakariosFast$e$freeInference2018,papamakariosSequentialNeuralLikelihood2019,greenbergAutomaticPosteriorTransformation2019,krouglovaMultifidelitySimulationbasedInference2025,thieleSimulationEfficientCosmologicalInference2025,saoulisTransferLearningMultifidelity2025}.

Beyond the challenge of learning the posterior, the curse of dimensionality also complicates validation of the inferred distributions. Coverage-based diagnostics such as TARP \citep[][]{lemosSamplingBasedAccuracyTesting2023} rely on comparisons of distances between posterior samples, reference draws, and ground truth parameters. As pairwise distances become increasingly similar in high dimensions, the discriminative power of these diagnostics is reduced. Moreover, coverage-based diagnostics provide only average-case guarantees rather than stringent tests near a fixed realization, as shown in an inference setting with a single parameter \citep{alokdaCoverageNotEnough2026}. All of the above motivates stress-testing posterior geometry conditioned on one realization in high-dimensional field-level inference.

\subsection{Training data variants}
To assess the impact of limited training data on inference conditioned on a fixed observation $\boldsymbol{y}_o$ (see section~\ref{sec:cosmo_case} for the simulator and setup used), we consider two limiting training-data regimes. Given access to samples from the target posterior in this study, these correspond to models trained on randomly sampled input--output pairs and models trained directly on posterior samples. In the random (\textit{rnd}) setting, models are trained on randomly generated joint samples. In the \textit{hmc} setting, we instead make use of samples from the posterior distribution $\pi_{\mathrm{HMC}}(\boldsymbol{x}\mid\boldsymbol{y}_o)$ obtained via HMC (section~\ref{sec:true_posterior}), which provide dense coverage of the posterior conditioned on the fixed observation $\boldsymbol{y}_o$. 

An overview of these training regimes is given in Table~\ref{tab:training_regimes}. Throughout most of this work, a maximum training dataset size of $59\,880$ samples is used to ensure a consistent comparison between architectures and training strategies, as well as against the reference posterior obtained from HMC. In addition, one model was trained on an extended dataset of $262\,144$ samples to assess convergence with increasing training data. This can be compared to the $\lesssim 2000$ simulations typically used in field-level SBI in cosmology \citep{List2023,Legin2023,Savchenko2024,Savchenko2025}, largely limited by the size of publicly available cosmological simulation suites \citep[e.g.][]{Villaescusa-Navarro2020}, and the $50\,000$ samples used in \citet{cuesta2024joint}. 

In this work, the training samples conditioned on $\boldsymbol{y}_o$ are used differently depending on the model class. For GLOW, the \textit{rnd} setting corresponds to conditional density estimation of $\pi(\boldsymbol{x}\mid\boldsymbol{y})$ with every $\boldsymbol{y}$ paired with one $\boldsymbol{x}$, while the \textit{hmc} setting reduces to unconditional density estimation of $\pi(\boldsymbol{x}\mid\boldsymbol{y}_o)$, i.e. with $\boldsymbol{y}=\boldsymbol{y}_o$ for all $\boldsymbol{x}$. For SI, the \textit{rnd} setting defines a transport from the data $\boldsymbol{y}$ to the corresponding initial conditions $\boldsymbol{x}$, whereas in the \textit{hmc} setting the interpolant transports from a narrow Gaussian distribution centered on $\boldsymbol{y}_o$ to samples from the posterior $\pi_{\mathrm{HMC}}(\boldsymbol{x}\mid\boldsymbol{y}_o)$.

\section{Field-Level SBI
in a Controlled Setting}
\label{sec:cosmo_case}

The true posterior is typically inaccessible and is precisely the quantity we aim to infer, making it difficult to assess how well a learned generative model captures the target distribution. To enable a principled evaluation and to allow training on the posterior samples, we adopt a controlled setting in which the posterior is available. Specifically, we use a first-order perturbative, physically meaningful forward model in the form of a simulator and observational noise, for which converged Hamiltonian Monte Carlo (HMC) produces high-quality posterior draws. The forward model and data generation procedure are described in section~\ref{sec:case_study}, while the HMC providing access to the posterior is introduced in section~\ref{sec:true_posterior}.

\subsection{Field-level cosmological inference}
\label{sec:case_study}

Field-level inference seeks to recover the underlying physical fields that give rise to observed data, together with reliable uncertainty estimates. In cosmology, this corresponds to reconstructing the cosmic initial conditions that evolved into the present-day large-scale structure of the Universe.  Both the initial conditions $\boldsymbol{x}$ and the observed data $\boldsymbol{y}$ are discretized three-dimensional fields. In our setting, each field lives on a $32^3$ grid, yielding a high dimension of $d = 32\,768$.

\subsubsection{Forward model}
\label{sec:forward_model}

A wide range of cosmological simulators of varying fidelity exists. To obtain a controlled setting with access to the ground-truth posterior, we adopt the structure formation model first-order LPT \citep{Zeldovich1970,bernardeauLargeScaleStructureUniverse2002}, denoted $\mathcal{S}$, and a boxsize of the cosmological volume of $1h^{-1}\,\mathrm{Gpc}$. The initial conditions are generated from a Gaussian white-noise field $\boldsymbol{x}$, which is mapped to primordial density fluctuations by applying the square root of the primordial power spectrum in Fourier space. The resulting field is then evolved forward using $\mathcal{S}$. We further add Gaussian observational noise $\boldsymbol{\varepsilon}$ to the simulation output, yielding the noise-contaminated matter density field used as data, i.e.,
\begin{equation}
    \boldsymbol{y}_o = \mathcal{S}(\boldsymbol{x}) + \boldsymbol{\varepsilon}, \qquad \boldsymbol{\varepsilon} \sim \mathcal{N}(\boldsymbol{0},\, \sigma^2 \mathbf{I}).
    \label{eq:mock_data}
\end{equation}
We choose $\sigma=0.3$ such that the signal dominates over the noise on all but the smallest scales (see Appendix~\ref{app:borg}).

\subsubsection{Data generation}
We generate the initial conditions field with $32^3$ grid elements by sampling from the white-noise prior $\pi(\boldsymbol{x}) = \mathcal{N}(\boldsymbol{0},\boldsymbol{I})$. We repeat this procedure $N=59\,880$ times, matching the number of independent samples obtained from HMC (see section~\ref{sec:true_posterior}). For each realization, we then run the simulator 
$\mathcal{S}$ to obtain the corresponding present-day (final) density field, also gridded to $32^3$. During training, we add noise to the simulation output on the fly, ensuring that
each pair $(\boldsymbol{x}, \boldsymbol{y})$ is sampled from the correct forward model
\begin{equation}
    (\boldsymbol{x}, \boldsymbol{y}) \sim  \pi(\boldsymbol{x})\pi_{\boldsymbol{\varepsilon}}\left(\boldsymbol{y} \mid \mathcal{S}(\boldsymbol{x})\right) ,
\end{equation}
with $\pi_{\boldsymbol{\varepsilon}}(\boldsymbol{y}\mid\mathcal{S}(\boldsymbol{x}))$ being the observational noise applied to the simulator output. This procedure effectively enriches the training distribution; for each underlying
simulation output, the model observes many realizations of $\boldsymbol{y}$ consistent with the same 
$\boldsymbol{x}$ given a fixed noise model. 

\subsection{Access to the true posterior as a controlled testbed}
\label{sec:true_posterior}
A key advantage of the forward model employed is its differentiability, which enables sampling from the exact posterior 
$\pi(\boldsymbol{x}\mid \boldsymbol{y}_o)$. We use the Hamiltonian Monte Carlo (HMC) sampler \citep{Duane1987,nealProbabilisticInferenceUsing1993} implemented in the \texttt{BORG} algorithm \citep{Jasche2012,Jasche2015,Lavaux2016,Jasche2019}, which infers the posterior distribution of cosmic initial conditions consistent with the observed data.  In this work, we apply it to a randomly simulated universe, where the present-day density field constitutes the data.

Although the posterior is not available in closed form, the combination of a Gaussian prior, a differentiable forward model, gridding, and Gaussian noise yields a well-behaved quasi-Gaussian posterior density. Under these conditions, the HMC results in an ergodic Markov chain whose stationary distribution is the exact posterior $\pi(\boldsymbol{x}\mid \boldsymbol{y}_o)$, allowing us to recover its geometry and covariance structure from long-run samples. In total, we generate $\sim 3\times10^6$ samples, each being a full $3$D field of $32^3$ grid elements, using three different Markov chains given the same $\boldsymbol{y}_o$. We provide convergence diagnostics and further details on the HMC chains in Appendix~\ref{app:borg}. We collect $59\,880$ samples that are approximately independent. In addition to being used as the reference posterior distribution, the set of samples $\boldsymbol{x} \curvearrowleft \pi_{\mathrm{HMC}}(\boldsymbol{x}\mid \boldsymbol{y}_o)$ is used as training data, allowing us to assess the extent to which this improves performance relative to training on random input–output pairs.

\begin{figure*}
\vskip 0.02in 
\begin{center}
\centerline{\includegraphics[width=0.99\textwidth]{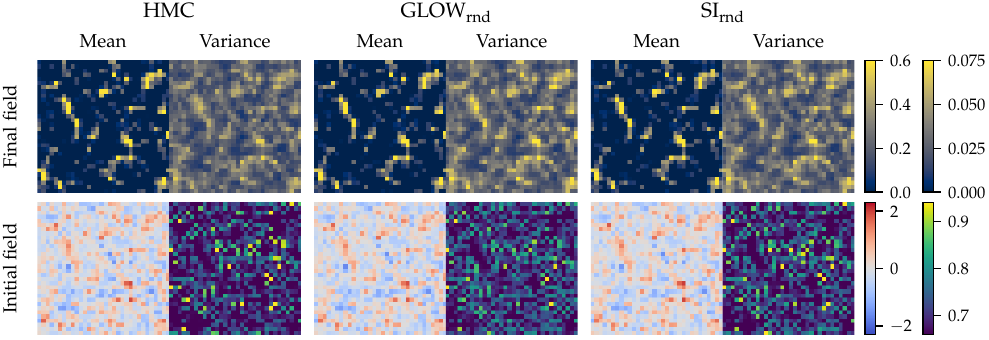}}
\caption{Voxel-wise posterior mean and variance in $2$D slices of the $3$D final present-day density field (top) and the initial conditions field (bottom) in a volume of $1h^{-1}$ Gpc and $\sim 30h^{-1}$ Mpc resolution per voxel. Results are estimated from samples drawn using HMC, a conditional GLOW model, and SI model trained on the \textit{rnd} data (see Table~\ref{tab:training_regimes}). Both GLOW and SI accurately recover the posterior mean and variance structure, though they tend to predict slightly broader posteriors.}
\label{fig:mean_var}
\end{center}
\vskip -0.2in
\end{figure*}

\begin{figure*}
\begin{center}
\centerline{\includegraphics[width=0.99\textwidth]{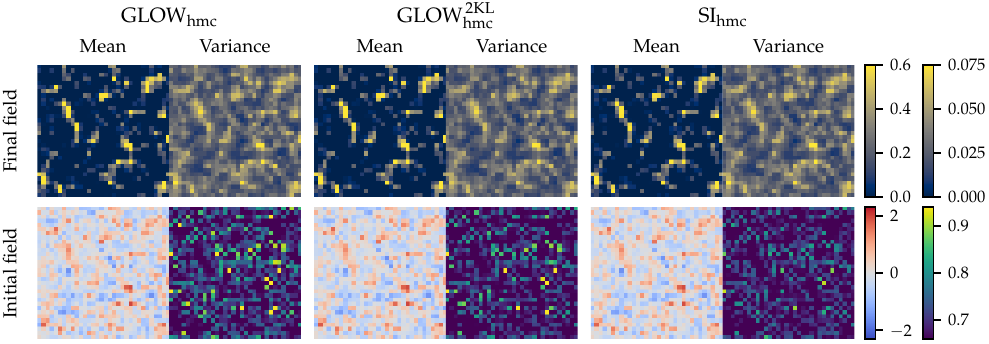}}
\caption{Same as Figure~\ref{fig:mean_var}, but for models trained on \textit{hmc} samples. Shown are GLOW trained with the forward KL, GLOW trained with both forward and reverse KL, and the SI. Training on \textit{hmc} samples leads to improved agreement in the posterior variance with HMC (see left-most column of Figure~\ref{fig:mean_var}), with the inclusion of the reverse KL term yielding visibly more accurate variance structure.}

\label{fig:hmc_trained_mean_var}
\end{center}
\vskip -0.2in
\end{figure*}

\section{Generative Models for Posterior Learning}
\label{sec:gen_models}

We introduce the generative models used for posterior learning over $3$D fields. Our goal is to assess how architectural choices, training data variants, and training objectives affect posterior approximation. To this end, we consider two complementary classes of models: explicit likelihood-based models and implicit generative models. We design both generative models to operate on $3$D fields and to respect the periodic boundary conditions of the cosmological simulation (details in Appendix~\ref{app:glow} and \ref{app:SI}). 

\subsection{Normalizing flows (GLOW)}

Normalizing flows model probability distributions through invertible transformations, enabling exact likelihood evaluation and efficient sampling \citep{dinhNICENonlinearIndependent2015, dinhDensityEstimationUsing2017}. We use the GLOW architecture \citep{kingmaGlowGenerativeFlow2018}, which is well suited for expressive modeling of $3$D fields. For posterior learning from simulated data, we employ a conditional variant of GLOW \citep{luStructuredOutputLearning2020a}, in which the flow is conditioned on the observed field $\boldsymbol{y}$ (see Appendix~\ref{app:glow}). 

To train our flows, we minimize a Kullback--Leibler (KL) divergence between the target distribution 
($\pi(\boldsymbol{x}\mid\boldsymbol{y})$ or $\pi(\boldsymbol{x}\mid\boldsymbol{y}_o)$), and the model distribution $q_\theta(\boldsymbol{x})$. The forward KL loss becomes (see Appendix~\ref{app:KL})
\begin{equation}
    \mathcal{L}_{\mathrm{fwdKL}} = 
- \frac{1}{N}\sum_{i=1}^N \log q_\theta(\boldsymbol{x}_i),
    \label{eq:fwdKL}
\end{equation}
and relies on samples drawn from the target distribution, i.e. $\boldsymbol{x}_i \sim \pi(\boldsymbol{x}\mid\boldsymbol{y})$ or $\boldsymbol{x}_i \sim \pi(\boldsymbol{x}\mid\boldsymbol{y}_o)$, which corresponds to the training data. We also train a model leveraging the reverse KL
\begin{equation}
    \mathcal{L}_{\mathrm{revKL}} = 
\frac{1}{N}\sum_{i=1}^N \log q_\theta(\boldsymbol{x}_i) - \log \pi (\boldsymbol{x}_i),
    \label{eq:revKL}
\end{equation}
which relies on samples generated by the GLOW model during training, $\boldsymbol{x}_i \sim q_\theta(\boldsymbol{x})$. In addition, in contrast to the forward KL objective, it requires evaluation of, and backpropagation through, the target density $\pi(\boldsymbol{x} \mid \boldsymbol{y})$. Since this density evaluation incorporates the forward simulator through the likelihood, $\pi(\boldsymbol{x}\mid\boldsymbol{y}) \propto \pi_{\boldsymbol{\varepsilon}}\left(\boldsymbol{y} \mid \mathcal{S}(\boldsymbol{x})\right)$, differentiable access to the simulator is required. To emulate the setting in which the simulator is non-differentiable, we replace the forward model $\mathcal{S}$ with a differentiable field-level neural emulator $\hat{\mathcal{S}}$ in the density evaluation (see Appendix~\ref{app:emulator} for implementation and validation of the emulator). Notably, using an emulator instead of the true simulator implies optimizing an approximate target density $\hat{\pi}(\boldsymbol{x} \mid \boldsymbol{y})$ rather than the exact density $\pi(\boldsymbol{x}\mid \boldsymbol{y})$. As shown in the results, this approximation is sufficient for the reverse KL term to improve performance.

We train models using $\mathcal{L}_{\mathrm{fwdKL}}$ on both the \textit{rnd} and \textit{hmc} data (Table~\ref{tab:training_regimes}). We further train a GLOW model using $\mathcal{L}=\mathcal{L}_{\mathrm{fwdKL}}+\lambda\mathcal{L}_{\mathrm{revKL}}$ with $\lambda=5$ on the \textit{hmc} samples.

\subsection{Stochastic Interpolants (SI)}

Stochastic interpolants provide a flexible framework for constructing diffusion processes that interpolate between arbitrary distributions \citep{Albergo2025}, generalizing standard diffusion models that evolve from a fixed Gaussian prior to a target distribution \citep{hoDenoisingDiffusionProbabilistic2020}. This formulation is particularly well-suited to large-scale structure inference as discussed in \citet{cuesta2024joint}, as the simulated data and the initial conditions share substantial overlap on linear, large scales. In this work, we additionally inject Gaussian noise into the simulation outputs, providing a simple step toward observational realism while retaining a controlled setting.

The generative process is defined through a continuous stochastic interpolation indexed by an interpolation parameter $s \in [0,1]$, whose dynamics are governed by a drift field. We parameterize this drift using a convolutional V-Net $\hat b_s$ and train it by minimizing the mean-squared deviation between the predicted drift and the instantaneous velocity $\boldsymbol{R}_s = \partial_s \boldsymbol{I}_s$ of the interpolant $\boldsymbol{I}_s$ along the stochastic path,
\begin{equation}
    \mathcal{L}_{\mathrm{drift}} = \frac{1}{N} \sum_{i=1}^{N}
\left|
\hat{b}_{s_i}(\boldsymbol{I}_{s_i}, \boldsymbol{y}, s_i)
- \boldsymbol{R}_{s_i}
\right|^2,
\quad s_i \sim \mathcal{U}(0,1)
    \label{eq:si_loss}
\end{equation}
such that the model inputs are the interpolant, the data $\boldsymbol{y}$, and $s$. This objective encourages the network to produce updates that move samples along the interpolation path between the reference and target distributions (see full details in Appendix~\ref{app:SI}). 
\section{Results}
\label{sec:results}

In this section, we present a set of diagnostics for a single realization of mock data, on which both the HMC chains and neural models are based. After training (details in Appendix~\ref{app:glow_training} and \ref{app:si_training}), we generate $59\,880$ samples from each model, matching the number of samples from the HMC reference posterior. For GLOW, posterior samples are obtained by sampling from the latent base distribution and applying the inverse flow. For SI, samples are generated by initializing at $\boldsymbol{y}=\boldsymbol{y}_o$ or by drawing from $\mathcal{N}(\boldsymbol{y}_o,\frac{1}{10}\mathbf{I})$ (see Table~\ref{tab:training_regimes}), and evolving the system using $100$ drift steps. Throughout, we refer to the sampled fields $\boldsymbol{x}$ as \emph{initial fields}, with the corresponding \emph{final fields} obtained by forward simulation via $\mathcal{S}$ (see Section~\ref{sec:forward_model}). 

To ensure that the conclusions drawn are not specific to a single realization, we repeat the analysis on two additional ground truth realizations (results presented in Appendix~\ref{app:diff_ground_truths}) and observe consistent behavior across all realizations.

\subsection{Voxelwise marginalized posteriors}
We begin by comparing voxelwise marginalized posterior statistics.
Fig.~\ref{fig:mean_var} shows $2$D slices of the voxelwise posterior mean and variance for the initial and final fields, using samples drawn from HMC, the conditional GLOW model, and the SI model when the generative models were trained on the \textit{rnd} set. Both models recover the posterior mean well, in close agreement with HMC. All methods reproduce the characteristic spatial structure of the HMC posterior uncertainty, though with an increased overall amplitude in the variance fields. Fig.~\ref{fig:hmc_trained_mean_var} shows the same quantities as Fig.~\ref{fig:mean_var}, but for models trained on \textit{hmc} samples. We see qualitatively that training on \textit{hmc} improves agreement in the posterior variance, with the inclusion of the reverse KL term yielding variance fields that most closely match the HMC reference. In Appendix~\ref{app:additional} (Fig.~\ref{fig:comp_all}), we provide the mean predictions compared with the ground truths for all models.

\begin{figure*}
\vskip 0.02in
\begin{center}
\centerline{\includegraphics[width=1.0\textwidth]{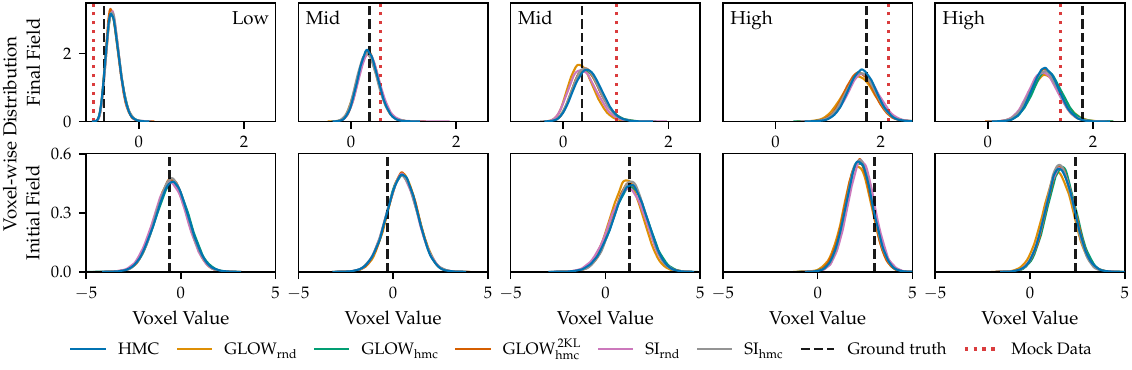}}
\caption{Marginalized posterior distributions for representative voxels spanning high-, intermediate-, and low-density environments. The mock data (dashed red) correspond to the true final field (dashed black) with added Gaussian noise. All methods yield consistent posteriors and typically recover values closer to the underlying truth, demonstrating that voxel-wise noise does not prevent them from exploiting large-scale correlations in the data. While these marginal distributions provide a useful local diagnostic, agreement at the voxel level alone is insufficient to validate the high-dimensional posterior geometry.}
\label{fig:voxel_distr}
\end{center}
\vskip -0.2in
\end{figure*}

\begin{figure*}
\centerline{\includegraphics[width=0.99\textwidth]{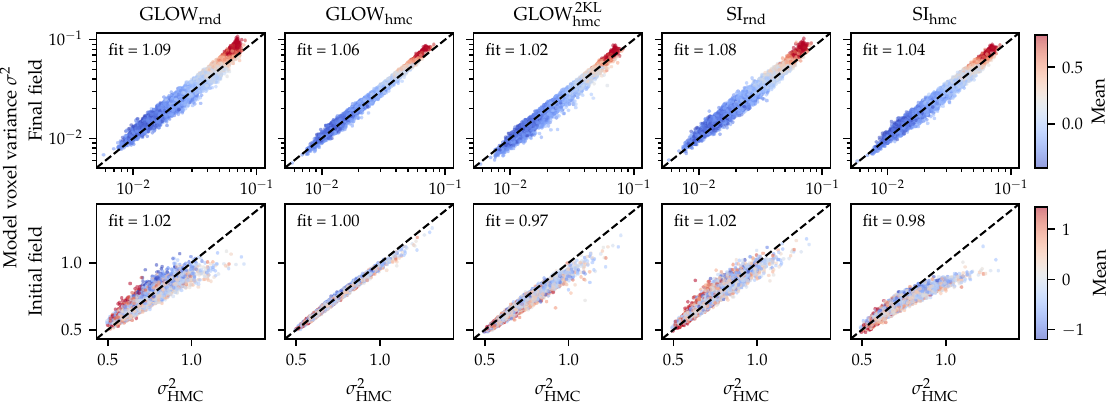}}
\caption{Voxel-wise posterior variance over samples of the reconstructed final (top) and initial (bottom) fields compared to the HMC posterior variance, where each point corresponds to a voxel. The dashed line indicates perfect agreement, while colours denote the posterior mean density, illustrating that the spatial structure of the variance of the final field closely follows the underlying density field. The reported fit values correspond to the median variance ratio, $\mathrm{median}(\sigma^2/\sigma^2_{\mathrm{HMC}})$. Training on \textit{hmc} improves agreement with the HMC variance field, particularly in the final field.}
\label{fig:variance_scatter}
\end{figure*}

\begin{figure}
\vskip 0.02in
\begin{center}
\centerline{\includegraphics[width=1.0\columnwidth]{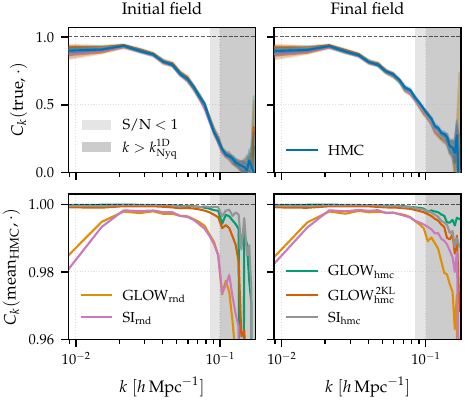}}
\caption{Scale-dependent cross-correlation between reconstructed and true fields for the initial (left) and final fields (right). The top row correlates samples from HMC, GLOW, and SI to the truth, while the bottom row shows mean fields of GLOW/SI relative to the HMC mean.}
\label{fig:crosscorr}
\end{center}
\vspace{-2em}
\end{figure}

\begin{figure}
\begin{center}
\centerline{\includegraphics[width=1.00\columnwidth]{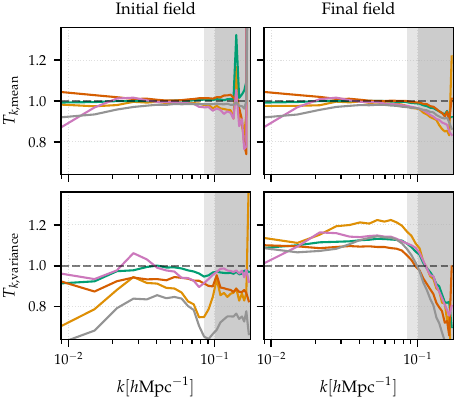}}
\caption{Power-spectrum ratios relative to HMC. Top row: the ratio for the posterior mean, which probes the recovery of spatial structure across scales. Bottom row: the ratio for the variance, quantifying spatial correlations in posterior uncertainty across scales. Colors match Fig.~\ref{fig:crosscorr}.}
\label{fig:powspec}
\end{center}
\vspace{-2em}
\end{figure}

We next examine marginalized posterior distributions by estimating voxel-wise kernel density estimates (KDEs) from the finite posterior samples. Fig.~\ref{fig:voxel_distr} shows five representative voxels spanning a range of density environments from low to high. In all cases, both GLOW and SI accurately reproduce the marginalized distributions, closely matching HMC and covering the ground truth. Notably, the posterior distributions are often centered closer to the underlying truth than the noisy data, reflecting the ability of the methods to exploit large-scale correlations in the data as they jointly reconstruct the full field, i.e. all voxels simultaneously. While these examples demonstrate good agreement at the level of individual marginals, a comprehensive assessment requires examining all voxels. This is shown in Fig.~\ref{fig:variance_scatter}, which compares the marginal voxel-wise posterior variances against HMC. Training on HMC samples systematically improves agreement with the HMC variance field in the final conditions for both GLOW and SI. For SI, however, this improvement is accompanied by a shift in the inferred initial-condition variances, highlighting that matching marginal uncertainties in the final field does not necessarily translate to identical uncertainty structure in the reconstructed initial conditions.

\subsection{Scale-dependent accuracy}
In the previous section, we assessed posterior accuracy using voxel-wise diagnostics on the inferred initial conditions and the corresponding posterior predictive final fields. Although these posterior resimulations can capture some coupling between spatial scales through the forward model, they remain voxel-wise statistics. To directly assess the recovery of structures across different spatial scales, we compute the scale-dependent cross-correlation between two fields $\boldsymbol{x}_a$ and $\boldsymbol{x}_b$, defined as

\begin{equation}
C_k(\boldsymbol{x}_a, \boldsymbol{x}_b)
=
\frac{P_k^{ab}}
{\sqrt{P_k^{aa}\,P_k^{bb}}},
\end{equation}
where the wavenumber $k$ is inversely proportional to physical scale. The auto- and cross-power spectra are estimated by averaging over $|\mathcal{K}(k)|$ Fourier modes within a shell $\mathcal{K}(k)$,
\begin{equation}
P_k^{ab}
=
\frac{1}{|\mathcal{K}(k)|}
\sum_{\boldsymbol{k}\in\mathcal{K}(k)}
\tilde{\boldsymbol{x}}_a(\boldsymbol{k})\,
\tilde{\boldsymbol{x}}_b^{*}(\boldsymbol{k}),
\label{eq:pow_def}
\end{equation}
with $\tilde{\boldsymbol{x}}(\boldsymbol{k})$ denoting the Fourier transform of the field. By construction, $C_k \in [-1,1]$, with values close to unity indicating strong phase alignment and accurate recovery of structure at the corresponding scale. We use the \texttt{PYLIANS3} package to compute these \citep{2018ascl.soft11008V}.

Figure~\ref{fig:crosscorr} shows the scale-dependent cross-correlation of reconstructed fields under two settings: correlations between individual posterior samples and the ground-truth fields (top), and correlations between posterior mean fields from the generative models and the HMC posterior mean (bottom). For individual samples, both generative models closely reproduce the HMC results across all scales for both the initial and final fields. The departure from unity arises from the presence of observational noise (see Fig.~\ref{fig:pk_noise} in Appendix~\ref{app:borg}). The correlations for the initial field begin to decrease at somewhat larger scales. This behavior can be understood in terms of gravitational collapse, which compresses matter into increasingly compact, overdense structures. Because we use the same finite grid resolution for both the initial and final fields, information about sufficiently small-scale fluctuations in the initial conditions is not retained in the gridded final density field. In the bottom row of Fig.~\ref{fig:crosscorr}, we compare the posterior mean fields to the HMC posterior mean for the initial conditions (left) and final fields (right). The \textit{rnd}-trained models show the largest discrepancies, exhibiting both suppressed small- and large-scale correlations. In contrast, training on \textit{hmc} samples substantially improves the agreement with the HMC posterior mean across a broad range of scales.

While cross-correlations probe spatial agreement of the reconstructed signal, they are largely insensitive to stochastic variations around the posterior mean. Instead, the diagnostic is dominated by the shared deterministic component of the reconstruction. Consequently, cross-correlation diagnostics are not sensitive to how uncertainty is distributed across the volume and can remain high even when the posterior variance is incorrect. To complement this, Fig.~\ref{fig:powspec} shows the transfer function, defined as the ratio of the auto-power spectra ($a=b$ in Eq.~\eqref{eq:pow_def}) relative to HMC,
\begin{equation}
T_k = \sqrt{P_k / P_{k,\mathrm{HMC}}}.
\end{equation}
As HMC accurately explores the correct power spectrum of the ground-truth fields (as visible in Fig.~\ref{fig:borg_warmup} in Appendix~\ref{app:borg}), this comparison provides insight into the degree to which the generative models recover both the mean signal as well as the uncertainty structure of the posterior. We note that the power spectrum ratio of the posterior mean characterizes the scales on which coherent structure is recovered, while the power spectrum ratio of the posterior variance probes the spatial correlation structure of the uncertainty and therefore higher-order aspects of the posterior geometry. While the mean-field power spectra agree with the ground truth to within $10\%$ across most scales, the voxelwise posterior variance shows deviations of up to $\sim30\%$. This shows that accurate recovery of the mean does not necessarily imply accurate recovery of the spatial organization of posterior uncertainty. Because the power spectrum of the variance field involves correlations of squared fluctuations, it is sensitive to higher-order moments of the posterior distribution, including fourth-order statistics. The larger discrepancies indicate that accurately reproducing posterior uncertainties is more challenging than recovering the mean structure alone. 

It is interesting to note that both the GLOW and SI models, despite differing in both architecture and training objective, exhibit similar offsets in the power spectrum ratio of the posterior variance fields relative to HMC. This may indicate limitations related to the available training data and/or the capacity of the neural generative models to accurately capture the posterior geometry. In both cases, however, using \textit{hmc} training data generally results in better performance. Notably, $\mathrm{SI}_{\mathrm{hmc}}$ achieves among the highest cross-correlations in Fig.~\ref{fig:crosscorr}, but shows slightly worse agreement in power-spectrum amplitude in Fig.~\ref{fig:powspec}. This is not necessarily contradictory, as the cross-correlation coefficient is insensitive to amplitude and primarily measures phase agreement, whereas the power spectrum is sensitive to amplitude fluctuations. These results may reflect a trade-off between optimizing the two objectives.

\begin{figure*}
\vskip 0.02in
\begin{center}
\centerline{\includegraphics[width=0.99\textwidth]{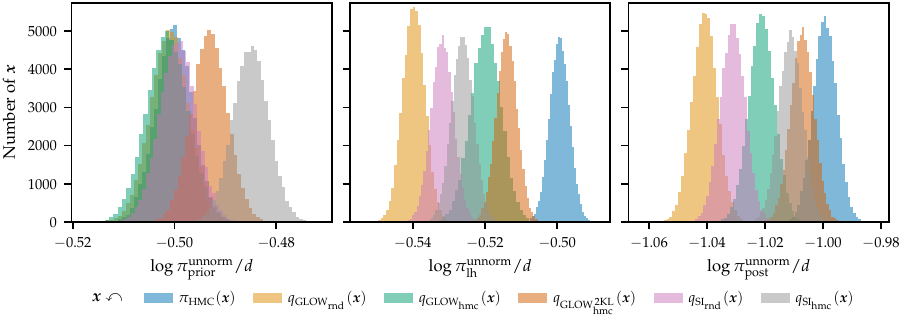}}
\vspace{-0.5em}
\caption{Histograms of unnormalized log prior, log likelihood, and log posterior evaluations, all divided by the dimension $d=32^3$ to aid visualization and interpretation. Thus, the values shown correspond to contributions per voxel (e.g. the Gaussian prior contributes $-1/2$ per dimension). Distributions of samples generated by each method reflect how effectively each sampler ultimately captures regions of the target posterior. The improved performance of GLOW$_{\mathrm{hmc}}^{\,2\mathrm{KL}}$ highlights
the impact of the training objective, and across models, training on \textit{hmc}
samples consistently yields better alignment with the target posterior
than training on randomly paired simulations.}
\label{fig:priorlhpost}
\end{center}
\vskip -0.3in
\end{figure*}

\subsection{Evaluation via target posterior}
\label{sec:target-posterior}

To directly assess whether the generative models reproduce the correct target distribution, we next evaluate samples under the explicit prior, likelihood, and posterior densities. For all likelihood evaluations, the generated initial condition samples are evolved using the true simulator $\mathcal{S}$. For each sample, we compute the log prior,
\begin{equation}
\log \pi_{\mathrm{prior}}(\boldsymbol{x})
=
-\frac{1}{2}\sum_i \boldsymbol{x}_i^2 + C_{\mathrm{prior}},
\label{eq:prior}
\end{equation}
the log likelihood,
\begin{equation}
\log \pi_{\mathrm{lh}}(\boldsymbol{y}_o|\boldsymbol{x})
=
-\frac{1}{2\sigma_{\mathrm{data}}^2}
\sum_j \bigl(\boldsymbol{y}_{o,j} - S(\boldsymbol{x})_j\bigr)^2 + C_{\mathrm{lh}},
\label{eq:lh}
\end{equation}
and the resulting log posterior,
\begin{equation}
\log \pi_{\mathrm{post}}(\boldsymbol{x|y}_o)
=
\log \pi_{\mathrm{prior}}(\boldsymbol{x})
+
\log \pi_{\mathrm{lh}}(\boldsymbol{x}) + C_{\mathrm{post}}.
\label{eq:post}
\end{equation}
As all additive constants are independent of $\boldsymbol{x}$ and identical across sampling methods, they do not affect comparisons between sample sets and are omitted here. Figure~\ref{fig:priorlhpost} shows that training the generative models on \textit{hmc} samples shifts the likelihood distributions closer to the HMC reference, indicating improved consistency with the data under the forward model. For GLOW, combining forward- and reverse-KL objectives further sharpens this agreement, bringing both likelihood and posterior distributions closer to HMC. Overall, both GLOW and SI generate high-quality samples that broadly reproduce the target prior, likelihood, and posterior distributions. However, neither achieves full overlap with the HMC reference, revealing residual discrepancies in the inferred posterior. Notably, efforts to improve posterior agreement, particularly for 
$\pi_{\mathrm{GLOW}_{\mathrm{hmc}}^{\,2\mathrm{KL}}}(\boldsymbol{x})$ and $\pi_{\mathrm{SI}_{\mathrm{hmc}}}(\boldsymbol{x})$, are accompanied by deviations in the prior term. In the absence of model misspecification, this is rather explained by the model identifying configurations of the initial field that yield a good fit to the mock data (final field plus noise) by slightly departing from Gaussianity.

\subsection{Log-likelihood ratios per sample}
We find that GLOW and SI achieve comparable performance in terms of sample quality and scale-dependent accuracy, with differences depending on training data and objective. In this section, we focus exclusively on GLOW models, since only these provide exact evaluations of probability densities by looking at pairs of samples. We test whether GLOW assigns correct \emph{relative posterior probabilities} to different realizations. In the conditional setting, absolute posterior densities cannot be compared due to the intractable normalization constant $\pi(\boldsymbol{y}_o)$. However, ratios of posterior probabilities between two samples conditioned on the same data remain accessible, as the normalization cancels. For a pair $(\boldsymbol{x}_a,\boldsymbol{x}_b)$, we therefore evaluate the true log posterior ratio
\begin{equation}
\mathrm{PR}
\equiv
\log \frac{\pi(\boldsymbol{x}_a \mid \boldsymbol{y}_o)}{\pi(\boldsymbol{x}_b \mid \boldsymbol{y}_o)}
=
\log \frac{\pi(\boldsymbol{y}_o \mid \boldsymbol{x}_a)\,\pi(\boldsymbol{x}_a)}
{\pi(\boldsymbol{y}_o \mid \boldsymbol{x}_b)\,\pi(\boldsymbol{x}_b)},
\end{equation}
where we used Bayes law. This ratio using Eq.~\eqref{eq:prior} and Eq.~\eqref{eq:lh}, can then be  compared to the ratio predicted by GLOW, which directly gives $q_{\theta}(\boldsymbol{x}_a \mid \boldsymbol{y}_o)$ and $q_{\theta}(\boldsymbol{x}_b \mid \boldsymbol{y}_o)$ (see Eq.~\eqref{eq:flow_dens_estimation}).

\begin{figure*}
\vskip 0.02in
\begin{center}
\centerline{\includegraphics[width=0.99\textwidth]{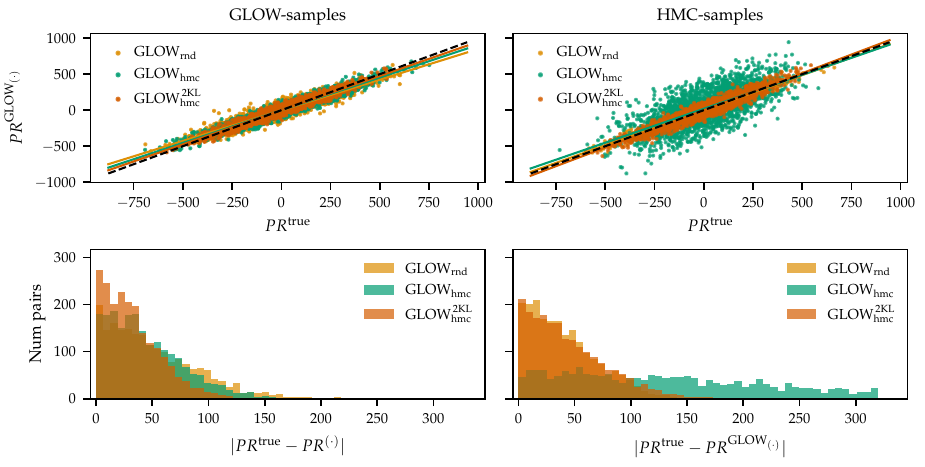}}
\caption{Posterior ratios (PR) of GLOW-predicted probabilities and true (Eq.~\eqref{eq:post}) log-posteriors for pairs of realizations conditioned on the same data. Each point corresponds to an independent random pair of samples, drawn either from GLOW (left) or from HMC (right). The corresponding lines are linear fits to each model. The bottom panels show the residuals from the dashed lines, which indicate perfect agreement, in the top panels. Training on HMC samples improves agreement, and combining forward- and reverse-KL objectives yields the best performance. GLOW yields probabilities that agree better for samples it generated itself compared to the HMC samples.}
\label{fig:loglikeglow}
\end{center}
\vskip -0.2in
\end{figure*}

Figure~\ref{fig:loglikeglow} shows that GLOW reproduces the qualitative ordering of posterior probabilities, but exhibits substantial quantitative discrepancies in log-probability space, corresponding to order-of-magnitude differences in probability. These discrepancies are more pronounced when evaluated on HMC samples than on GLOW-generated samples, indicating that the model assigns internally more consistent probabilities within the region of configuration space it explores, while struggling to accurately represent the broader posterior landscape. This behavior is reflected in the slopes shown in Figure~\ref{fig:loglikeglow}, obtained from linear fits to the predicted log-likelihoods, as well as increased scatter when evaluated on HMC samples. When evaluated on GLOW-generated samples, the \textit{rnd} model shows the largest deviation from perfect agreement between predicted and reference values, followed by the \textit{hmc} model, while the model trained on \textit{hmc} samples with both forward and reverse KL losses achieves the closest agreement. Overall, these results highlight the difficulty of accurately learning relative probabilities in high-dimensional field-level inference problems.

\section{Discussion}
\label{sec:discussion}
Having presented the results, we now discuss their implications and place them in a broader context. In particular, we consider the relevance of these findings for field-level SBI with more sophisticated forward models, the reliability of neural posterior approximations, and the amount of training data required even in this simplified setting. We also comment on possible methodological extensions and emerging neural models that may help address the limitations identified here.

\subsection{High-fidelity simulators}
In this work, we have focused on a computationally efficient forward model based on first-order Lagrangian perturbation theory. While this approximation is well suited to describe large-scale structure on linear scales, it does not capture the full non-linear dynamics of structure formation. However, already in this setting, we find that the considered generative models do not fully reproduce the posterior geometry obtained via HMC, highlighting that accurately learning high-dimensional posterior distributions remains challenging even under idealized conditions. At the same time, the present setup does not allow us to conclusively determine how such discrepancies would propagate to more realistic inference settings involving increasingly non-linear structure formation. The deviations identified here should therefore not be interpreted as evidence that neural generative approaches yield conservative posterior approximations. Rather, the results indicate that posterior reliability in high-dimensional field-level inference warrants further investigation in progressively more realistic settings.

Across all experiments, we find a consistent pattern. Both GLOW and SI recover posterior means, cross-correlations, and marginalized voxel distributions to percent-level accuracy. Performance further improves when training on HMC samples, which provide a denser representation of the target posterior than randomly drawn simulations. More sensitive diagnostics reveal larger discrepancies across all trained models. Power spectra of the posterior variance field deviate by up to $30\%$, evaluations under the explicit target posterior expose mismatches in the prior, likelihood, and posterior, and log-posterior ratio tests show that GLOW assigns incorrect relative probabilities to samples by orders of magnitude. Together, these results indicate that while generative models reproduce local and low-order statistics well, they struggle to recover the full geometry of the target posterior.

\begin{figure*}
        \centering
    \includegraphics[width=1.\linewidth]{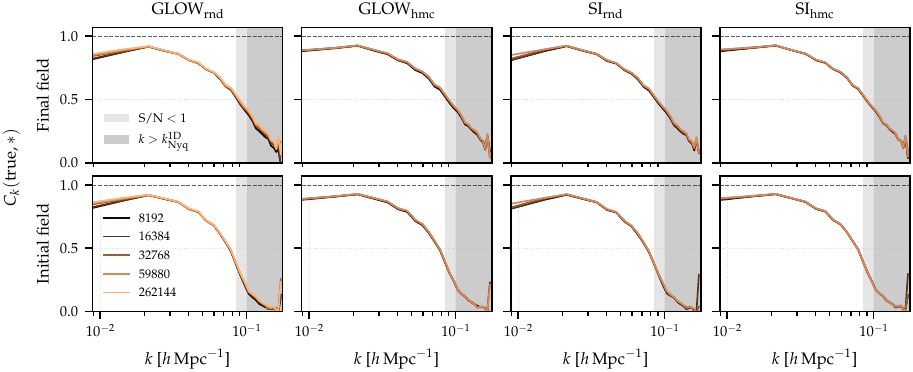}
    \caption{
Scale-dependent cross-correlation between reconstructed and true fields as a function of the number of training simulations for the GLOW and SI models. The top row shows the final fields and the bottom row the initial conditions. Increasing the number of training simulations yields modest improvements in the recovery of structures across scales, an effect that is slightly more evident in the final field. For $\mathrm{GLOW}_{\mathrm{rnd}}$, the training set is additionally extended to $262\,144$ simulations to probe convergence beyond the primary comparison regime of $59\,880$ samples. Since the cross-correlation primarily probes the posterior mean, convergence of these curves does not necessarily imply full posterior convergence, motivating the likelihood- and posterior-based diagnostics shown in Fig.~\ref{fig:num_of_sims}.
}
\label{fig:num_of_sims_cross}
\end{figure*}

Moving to more realistic non-linear forward models, such as efficient quasi-$N$-body approaches or full $N$-body simulations of dark matter with or without baryonic physics \citep[see][for a review of various methods]{Vogelsberger2020} are required for applications of field-level inference to real data. However, non-linear high-fidelity models introduces stronger mode coupling and non-Gaussian features in the resulting fields. The deviations identified in this work raise the question of how generative posterior approximations behave in increasingly non-linear inference settings, where posterior structure may become more complex. Addressing this systematically, including comparisons across forward models, is left for future work.

Additionally, increasing the fidelity of forward models may result in simulations that are no longer differentiable, limiting the applicability of classical gradient-based inference approaches. This is precisely the setting in which neural network–based methods have shown considerable promise, with recent work demonstrating that it is possible to navigate the space of cosmic initial conditions using data generated by non-differentiable forward models. This includes simulation-based approaches, such as field-level SBI \citep[e.g.][]{List2023,Legin2023,cuesta2024joint,Savchenko2024,Savchenko2025,bayerFieldLevelInferenceGalaxies2026}, as well as optimization-based methods \citep{Doeser2025}.

This, however, raises a key question: to what extent can posterior distributions obtained with neural network approaches remain scientifically reliable when exact reference methods and/or guarantees are no longer available? In particular, the absence of ground truth posteriors makes it challenging to assess the accuracy of learned approximations. This highlights an important trade-off. Posterior inference methods with convergence guarantees in the limit of infinitely long Markov chains, such as HMC, offer a principled framework for uncertainty quantification. While this asymptotic limit is never attained in practice, a range of established diagnostics can be used to assess sampling quality and convergence. However, such methods require gradients of the simulator and therefore often rely on approximate, differentiable forward models, potentially leading to an increased mismatch with the true data-generating process. Conversely, reducing model misspecification through more realistic forward models often introduces non-differentiable components that prevent the use of gradient-based sampling. In such settings, neural posterior approximations offer a promising alternative. These methods also possess asymptotic guarantees in the limits of infinite training data, sufficient model capacity, and an appropriate training objective. However, assessing how closely these conditions are satisfied in practice, and consequently the accuracy of the resulting posterior distributions, is generally more challenging. While increasing the physical fidelity of the forward model within field-level inference is a necessary direction from a modelling perspective, the resulting neural posterior distributions should therefore be interpreted with care, and systematic validation, including convergence studies with respect to the number of training simulations, remains essential.

\subsection{Importance of training data}
\label{sec:training_data_convergence}
The results obtained in section~\ref{sec:results} indicate that training on HMC samples and refining the learning objective tend to improve agreement with the target posterior, but do not fully eliminate discrepancies. In particular, training GLOW with a combination of forward and reverse KL divergences yields the best performance in several diagnostics among the tested variants, highlighting the sensitivity of posterior learning to both the training objective and the training data.

To assess the role of training data, we study how performance scales with the number of simulations used during training. We begin in Fig.~\ref{fig:num_of_sims_cross} by showing the scale-dependent cross-correlation between reconstructed and true fields as a function of training set size, following a similar convergence analysis to that of \citet{cuesta2024joint}. The main comparison between architectures and training strategies is performed using training sets of up to $59\,880$ samples. To probe convergence beyond this regime, we additionally train a $\mathrm{GLOW}_{\mathrm{rnd}}$ model on an extended dataset of $262\,144$ samples. We find that increasing the number of training simulations generally improves the recovery of structures across scales, with the trend being more apparent in the final fields. This highlights the value of posterior predictive diagnostics, which can reveal improvements that are less evident when considering the inferred initial conditions alone. However, cross-correlation statistics are dominated by the mean-field component and are largely insensitive to posterior uncertainties. Consequently, convergence of cross-correlations alone should not be interpreted as evidence of full posterior convergence. 

\begin{figure}
    \centering
    \includegraphics[width=1.\linewidth]{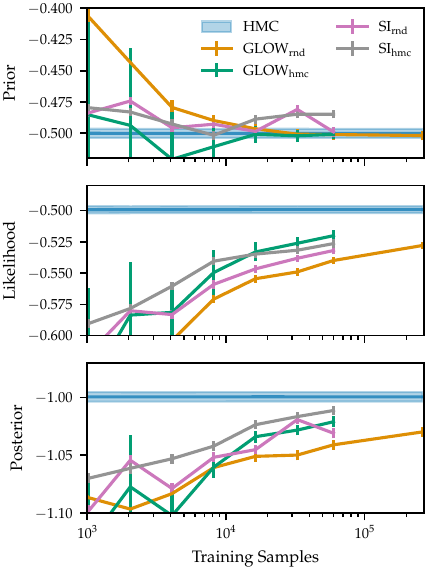}
    \caption{Performance as a function of the number of training samples, evaluated through the unnormalized log-prior (top), log-likelihood (middle), and log-posterior (bottom). Increasing the training set size systematically improves agreement with the HMC reference posterior for all models. The overall convergence behaviour appears similar across architectures and training strategies. Only $\mathrm{GLOW}_{\mathrm{rnd}}$
	was trained on the largest dataset ($262\,144$ simulations); its continued improvement suggests that convergence to the target posterior would require at least another order of magnitude more training data. In contrast, concentrating training samples in regions of high posterior density, as approximated here by the \textit{hmc} training sets, yields comparable performance with roughly an order of magnitude fewer simulations.}
    \label{fig:num_of_sims}
\end{figure}

Figure~\ref{fig:num_of_sims} shows the unnormalized log-prior, log-likelihood, and log-posterior diagnostics introduced in Fig.~\ref{fig:priorlhpost} as a function of training set size. Because these diagnostics directly probe the target posterior density, they provide a complementary measure of posterior accuracy. Unlike the cross-correlation coefficient, which compresses information from many Fourier modes within each $k$-bin, these metrics evaluate agreement in the full high-dimensional field. As a result, residual errors accumulate across all voxels, making them more sensitive to inaccuracies in the reconstructed realization. All methods show systematic improvement with increasing training set size, indicating that convergence has not yet been reached within the explored training regime. Even after extending the $\mathrm{GLOW}_{\mathrm{rnd}}$ training set to $262\,144$ samples, agreement with the reference posterior continues to improve without clear signs of saturation. This suggests that substantially larger training datasets, likely at least another order of magnitude, may be required to fully converge toward the HMC posterior. 

Training data from the \textit{hmc} training sets converge more rapidly and achieve comparable posterior accuracy with roughly an order of magnitude fewer simulations than the corresponding \textit{rnd} training sets. The \textit{hmc} training set thus provide an idealized demonstration of the gains obtainable from concentrating simulations in the most relevant regions of parameter space. Given the already substantial training requirements observed here, which may increase further for more realistic simulators, simulation-efficient approaches such as active learning, multi-fidelity strategies, and transfer learning are likely to play an important role \citep[e.g.][]{thieleSimulationEfficientCosmologicalInference2025,saoulisTransferLearningMultifidelity2025}.

\subsection{Rapid development of generative models}
While this work focuses on specific architectures, the field of generative modeling is evolving rapidly, with increasingly expressive and scalable variants being developed. This suggests that some of the limitations observed here may be alleviated by future model improvements. For example, recent advances in flow-based models have demonstrated strong performance in high-dimensional settings \citep{zhaiNormalizingFlowsAre2025,zhangFractalFlowHierarchical2025}. Exploring more expressive flow architectures in the context of field-level SBI, as well as extending these approaches to $3$D, is therefore a promising direction for future work, building on the performance established here for GLOW.

Furthermore, score-based diffusion models \citep{songScoreBasedGenerativeModeling2021} as used in \citet{Legin2023} for initial conditions sampling, and variational diffusion models \citep{kingmaVariationalDiffusionModels2023} used for sampling the underlying matter field from various cosmic tracers \citep{onoDebiasingDiffusionProbabilistic2024,parkProbabilisticReconstructionDark2023,chenFieldlevelReconstructionForegroundContaminated2025} constitute another class of generative models not studied in this work. We note, however, that Stochastic Interpolants constitute a recent unifying framework bridging flow-based and diffusion-based approaches \citep{Albergo2025}. The SI model employed in this work is thus related at the architectural level to the diffusion-based approaches mentioned above, and also relies on similar neural network backbones, such as U-Nets \citep{ronnebergerUNetConvolutionalNetworks2015}. One difference is that our SI model is trained via a flow-matching objective on the drift field, unlike score-based learning. This may lead to different empirical behavior, and a systematic comparison is left for future work.

While future improvements in model architectures may alleviate some of the limitations identified here, the findings of this work highlight the importance of rigorous posterior validation in high-dimensional generative inference. In particular, the diagnostics, validation strategies, and analyses of training objectives and training data regimes explored here can help guide the development and evaluation of future generative approaches toward the level of reliability required for scientific inference.

\section{Conclusion}
\label{sec:conclusion}

Generative models are increasingly used for high-dimensional scientific inference, in particular within simulation-based inference (SBI). In these settings, uncertainty quantification is essential, yet remains challenging due to the high dimensionality of the inferred quantities. As a result, evaluation is often based on low-order diagnostics, such as marginalized distributions, posterior means, or cross-correlations. While these statistics are informative and necessary, they provide only partial constraints on the full posterior structure. 

In cosmology, the problem of inferring the initial conditions of the Universe has seen growing use of deep learning–based approaches. Owing to the complexity of the underlying simulators, traditional MCMC methods are computationally demanding and often inapplicable in the presence of non-differentiable components. To enable full posterior inference with such simulators, generative models have recently been applied within field-level SBI \citep[][]{List2023,Legin2023,cuesta2024joint,Savchenko2024,Savchenko2025,bayerFieldLevelInferenceGalaxies2026}. In this work, we build on these developments by introducing a controlled field-level inference setting in which the target posterior is accessible via HMC. This allows us to systematically assess how well generative models—specifically GLOW and Stochastic Interpolants (SI)—reproduce the underlying posterior geometry. Our goal is to understand what current generative approaches capture, and where they fall short, when evaluated against a known reference posterior.

Taken together, our results demonstrate that reproducing mean statistics or low-dimensional marginals is not sufficient to guarantee posterior fidelity. This highlights the importance of complementary diagnostics. Throughout this work, we have applied analogous validation metrics on the posterior predictive distribution by propagating posterior samples through the forward simulator and analyzing the resulting observable fields. Such posterior predictive tests provide a stringent validation of posterior accuracy and should form an important component of the assessment of generative models for scientific inference. While generative models can yield visually convincing and statistically consistent results under common diagnostics, they may still exhibit deviations in aspects of the full posterior structure. In this work, we observe deviations in, for example, the power spectrum of the variance field and the voxel-based sample tests compared to HMC. Furthermore, although all methods exhibit systematic improvement with increasing training set size, convergence in the cross-correlation alone is not sufficient to guarantee convergence of the full posterior. Even in the controlled setting considered here, with an efficient differentiable simulator and access to reference HMC samples, our results suggest that substantially larger training datasets may be required to fully reproduce the target posterior.

At the same time, a more optimistic perspective is also warranted. These methods enable inference with complex simulators that would otherwise be computationally prohibitive or intractable with traditional approaches. The level of agreement achieved across selected diagnostics indicates that meaningful structure is already being captured. This, in turn, underscores the importance of more comprehensive evaluation strategies and careful interpretation of learned posterior distributions. Looking forward, several directions merit further investigation. First, understanding how these methods scale with increasing dimensionality and complexity of the forward model is critical for applications to larger and more complex datasets. Second, extending the comparison between SBI-based posteriors and traditional sampling approaches, such as HMC, to joint inference of cosmological parameters and initial conditions is a natural and important next step.

In conclusion, our controlled study, enabled by access to a reference posterior, allows for a detailed assessment of generative models in field-level SBI. We demonstrate both the promise and current limitations of neural generative approaches in capturing high-dimensional posterior structure. These findings highlight the need for more robust inference methods and more comprehensive evaluation practices for uncertainty quantification in high-dimensional scientific problems.
\section*{Data Availability}
The data underlying this article are available from the corresponding author upon request.

\section*{Acknowledgements}
We thank Adrian E. Bayer,  Carolina Cuesta-Lazaro, Matthew Ho, Guilhem Lavaux, Francisco Maion, Stuart McAlpine, Francisco Villaescusa-Navarro, and Benjamin D. Wandelt for useful discussions related to this work. In particular, we thank Adrian E. Bayer and Matthew Ho for their feedback on the manuscript.

This research utilized the Sunrise HPC facility supported by the Technical Division at the Department of Physics, Stockholm University. LD and JJ acknowledge support from the Simons Foundation
through the Simons Collaboration on "Learning the Universe". This
work was made possible by the research project grant "Understanding
the Dynamic Universe," funded by the Knut and Alice Wallenberg
Foundation (Dnr KAW 2018.0067). Additionally, JJ acknowledges
financial support from the Swedish Research Council (VR) through
the project "Deciphering the Dynamics of Cosmic Structure" (2020-
05143). 

\bibliographystyle{mnras}
\bibliography{GenMod4Science}

\appendix
\clearpage
\section{HMC and \texttt{BORG}}
\label{app:borg}

To establish a controlled reference for posterior inference with theoretical guarantees, we generate samples from the target posterior using Hamiltonian Monte Carlo (HMC), which provides asymptotic convergence guarantees and reversible Markov dynamics. We perform Bayesian inference of the $3$D gridded initial conditions of size $32^3$ voxels in a cosmological volume of $1h^{-1}$ Gpc using the \texttt{BORG} framework \citep{Jasche2012} with a first-order Lagrangian perturbation theory (1LPT) structure formation model, denoted $\mathcal{S}$. We use the cosmological parameters  $\Omega_\mathrm{m} = 0.3175$, $\Omega_\mathrm{b} = 0.049$, $h = 0.6711$, $n_\mathrm{s} = 0.9624$, $\sigma_8 = 0.834$, and $w=-1$.

\subsection{Ground truth data generation}
\begin{figure}
\vskip 0.02in
\begin{center}
\centerline{\includegraphics[width=\columnwidth]{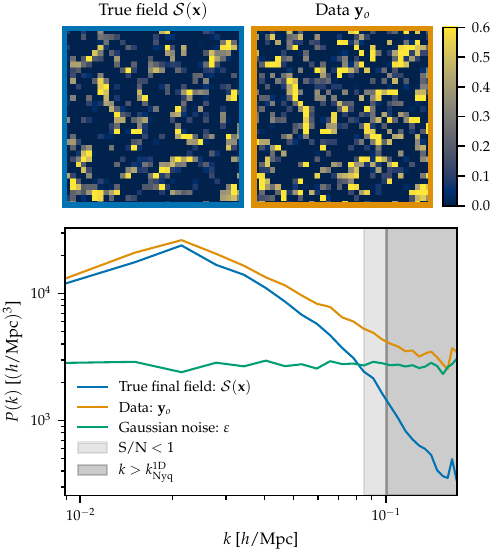}}
\caption{\textbf{Top:} $2$D slices through the $3$D true final density field $\mathcal{S}(\boldsymbol{x})$ and the corresponding mock data $\boldsymbol{y}_o$. The side-length is $1h^{-1}$ Gpc and the resoluton is $31.25h^{-1}$ Mpc. \textbf{Bottom:} Power spectra of the true final field $\mathcal{S}(\boldsymbol{x})$, the noisy data $\boldsymbol{y}_o$, and the Gaussian noise realization $\boldsymbol{\varepsilon}$. The shaded region indicates scales where the signal-to-noise ratio drops below unity. For $\boldsymbol{\varepsilon} \curvearrowleft \mathcal{N}(\boldsymbol{0},\sigma^2\mathbf{I})$ with $\sigma=0.3$, the signal dominates over the noise on most scales, with noise becoming comparable only on the smallest scales.}
\label{fig:pk_noise}
\end{center}
\vskip -0.2in
\end{figure}

Using Eq.~\eqref{eq:mock_data}, we generate mock observations from a fixed ground-truth realization of the initial conditions. We apply the simulator $\mathcal{S}$ to the initial conditions and add Gaussian noise to the simulation output, yielding a tractable likelihood. We choose the noise level such that the signal-to-noise ratio exceeds unity on most scales (Fig.~\ref{fig:pk_noise}).

\subsection{HMC sampling} 
\begin{figure*}
\vskip 0.02in
\begin{center}
\centerline{\includegraphics[width=0.99\textwidth]{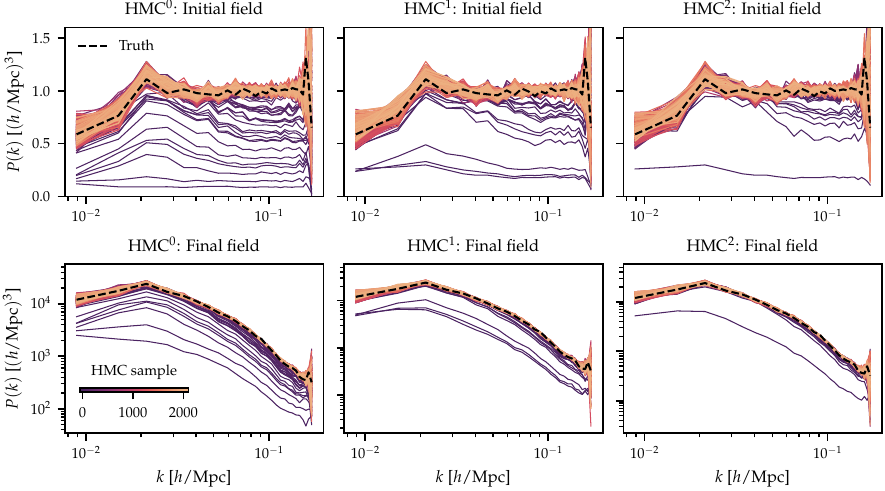}}
\caption{Evolution of three independent HMC chains. We monitor the power spectra of the inferred initial density fields (top), and the corresponding forward-simulated final fields (bottom). We show the first $2000$ samples, which constitute a conservative warm-up phase; every fifth sample is shown. The ground-truth fields are indicated by black dashed curves. The chains follow distinct early trajectories, but they all converge to sample around the target posterior consistent with the ground truth.}
\label{fig:borg_warmup}
\end{center}
\vskip -0.2in
\end{figure*}

Posterior sampling of the initial conditions field is carried out using Hamiltonian Monte Carlo (HMC). We run three independent Markov chains, each generating $10^6$ samples (each a $32^3$ cube) of the initial conditions. For each chain, we start from a downscaled (by a factor of $0.3$) random initial conditions field. Empirically, the warm-up phase is reached after $\sim 200$ samples, after which the chains enter the stationary regime. We monitor the warm-up phase of the Markov chain through the evolution of the power spectrum in Fig.~\ref{fig:borg_warmup}.

\subsection{Convergence}
\begin{figure}
\vskip 0.02in
\begin{center}
\centerline{\includegraphics[width=\columnwidth]{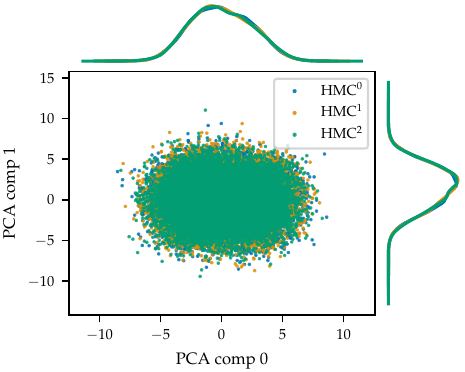}}
\caption{Scatter and marginal KDEs of three HMC chains projected onto the first two principal components of the pooled samples show agreement in the dominant posterior variance directions.}
\label{fig:pca}
\end{center}
\vskip -0.2in
\end{figure}

We adopt a conservative warm-up of $2000$ samples to provide a safety margin beyond the initial stabilization of the power spectrum. We then compute the Gelman--Rubin statistic across chains both for the negative log-likelihood of the forward-simulated density field with respect to the data, as well as voxel-wise for the inferred initial conditions. In all cases, we obtain $|R-1| < 10^{-4}$, indicating that the chains have converged and are sampling from the same target posterior. As an additional diagnostic, we perform a principal component analysis (PCA) of the sampled initial conditions on the joint basis to verify that all chains explore the same regions of parameter space. We project the samples onto the first two principal components and compare their joint distributions across chains, finding consistent coverage, as shown in Fig.~\ref{fig:pca}.

\subsection{Independent samples}
\begin{figure*}
\vskip 0.02in
\begin{center}
\centerline{\includegraphics[width=1.0\linewidth]{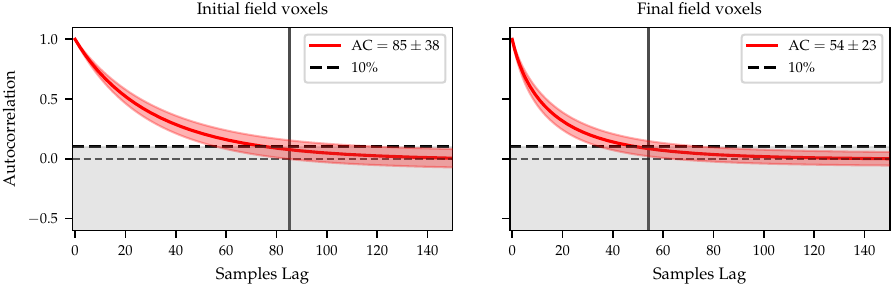}}
\caption{Autocorrelation functions of individual voxels in the initial (left) and final (right) fields, averaged over voxels. Vertical lines indicate the lag at which the autocorrelation falls below the $10\%$ level, a commonly used reference for approximate independence. We adopt a thinning of $50$ HMC iterations to sample the posterior, retaining weak correlations while providing a larger number of posterior samples.}

\label{fig:borg_acf}
\end{center}
\vskip -0.2in
\end{figure*}

To estimate the effective number of samples, we compute the autocorrelation function of both the inferred initial conditions and the corresponding forward-simulated final density fields, shown in Fig.~\ref{fig:borg_acf}. Based on the decay of the autocorrelation, we adopt a correlation length of $50$ samples. After discarding an initial the warm-up phase of $2000$ samples, we subsample each chain at this interval. This procedure yields a total of $59\,880$ samples, all of which are used in the subsequent analyses.

\section{GLOW}
\label{app:glow}
We adopt the GLOW architecture \citep{kingmaGlowGenerativeFlow2018}, which builds on the NICE and RealNVP flows \citep{dinhNICENonlinearIndependent2015,dinhDensityEstimationUsing2017} and has proven effective for high-dimensional image-like data. GLOW consists of a sequence of flow steps, each comprising actnorm, an invertible $1\times1$ convolution, and an affine coupling layer, arranged within a multi-scale architecture that captures structure across resolutions. We use $L=3$ levels and $K=8$ flow steps per level for all our GLOW models. For the convolutional neural networks in the affine coupling layers we use $64$ hidden channels. Furthermore, we employ a conditional variant of GLOW \citep{luStructuredOutputLearning2020a}, in which the flow is conditioned on the observed field $\boldsymbol{y}_o$. As cosmological simulations are defined with periodic boundary conditions, we employ periodic padding of both the conditional input fields and internal feature maps prior to convolutional operations in GLOW. While periodic padding increases the memory footprint during training and sampling, it eliminates boundary artifacts that would otherwise bias the learned posterior structure. In the following, we detail the GLOW architecture, as well as the training procedure and loss functions used.

\subsection{Change-of-variables formula and log-likelihood}

A flow defines an invertible transformation 
$f_\theta : \boldsymbol{z} \mapsto \boldsymbol{x}$, where 
$\boldsymbol{z} \sim \mathcal{N}(\boldsymbol{0},\mathbf{I})$ is a base variable. The conditional density is given by the change-of-variables formula,
\begin{equation}
\log q_\theta(\boldsymbol{x} \mid \boldsymbol{y})
  = \log p_0(\boldsymbol{z})
  + \sum_{\ell=1}^L \log \left| \det J_\ell \right|,
\label{eq:flow_dens_estimation}
\end{equation}
where $J_\ell$ is the Jacobian of the $\ell$-th invertible transformation. Training corresponds to maximizing this log-density over simulated  pairs $(\boldsymbol{x},\boldsymbol{y})$, which is equivalent to minimizing the forward KL divergence (see Appendix~\ref{app:KL}).

\subsection{Conditional GLOW}
We take inspiration from a conditional extension of GLOW \citep{luStructuredOutputLearning2020a}, in which all flow layers are conditioned on the observed field $\boldsymbol{y}$. In our model, conditioning is applied exclusively at the level of the affine coupling layers, which we found to provide stable training; more extensive conditioning of additional flow components could further increase expressivity, but is left for future work.

To avoid confusion with the physical initial and final fields used throughout this work, intermediate flow states are here denoted by $\tilde{\boldsymbol{x}}$ and $\tilde{\boldsymbol{y}}$. Each coupling layer splits the channel dimension of $\tilde{\boldsymbol{x}}$ into two parts, $\tilde{\boldsymbol{x}}_a$ and $\tilde{\boldsymbol{x}}_b$, and applies an affine transformation to one part while leaving the other unchanged:
\begin{equation}
\begin{split}
\tilde{\boldsymbol{y}}_a &= \tilde{\boldsymbol{x}}_a, \\
\tilde{\boldsymbol{y}}_b &=
\tilde{\boldsymbol{x}}_b \odot \exp\!\left( s_\theta(\tilde{\boldsymbol{x}}_b,\, g_\theta(\boldsymbol{y})) \right)
+ t_\theta(\tilde{\boldsymbol{x}}_b,\, g_\theta(\boldsymbol{y})) ,
\end{split}
\end{equation}
Here $(\boldsymbol{s}_\theta, \boldsymbol{t}_\theta)
= \mathrm{NN}_\theta\!\bigl(\tilde{\boldsymbol{x}}_b,\, g_\theta(\boldsymbol{y})\bigr)$ denote the scale and shift outputs of a single 3D convolutional neural network. The conditioning features $g_\theta(\boldsymbol{y})$ are produced by an additional 3D convolutional neural network applied to the data $\boldsymbol{y}$ and concatenated with $\tilde{\boldsymbol{x}}_b$ before predicting the affine parameters.  Even in the unconditional training, using the \textit{hmc} data, we retain the conditioning network, which is then always evaluated on the fixed observed field $\boldsymbol{y}_o$; we choose this design since it allows the model to utilize information from $\boldsymbol{y}_o$ when learning the posterior. The Jacobian of this affine transformation remains triangular, enabling efficient and exact computation of the log-determinant.

\subsection{Training}
\label{app:glow_training}
During model development, a held-out validation set was used to verify stable training, the absence of boundary artifacts, and the ability to reproduce correct spatial structures consistent with ground-truth simulations and HMC samples. For the final experiments, we train the model on all available samples for $150$ epochs with a fixed learning rate of $10^{-4}$. A separate validation set is not required at this stage, as model performance is assessed through downstream sample-based diagnostics rather than validation loss. We observe continued improvement in sample quality across the tests presented in section~\ref{sec:results} up to epoch $150$.

Memory usage is dominated by the model parameters and the chosen batch size, as all operations act on $3$D fields. The conditional model comprises approximately $22$M parameters. While the target fields in this work have a resolution of $32^3$, the conditioning field is padded by $8$ voxels in each direction, resulting in a field of size $48^3$. In addition, intermediate computations involve periodic padding by one or two voxels per dimension to mitigate boundary effects. Training is performed using TensorFlow’s mirrored strategy with a global batch size of $64$. Each batch is distributed across four NVIDIA A100 GPUs with $40\,\mathrm{GiB}$ of memory, processing $16$ samples per GPU.

\subsection{Loss functions: Kullback--Leibler Divergence}
\label{app:KL}
Normalizing flows are typically trained by minimizing a Kullback--Leibler (KL) divergence between a target distribution $\pi$ and the model distribution $q_\theta(\boldsymbol{x})$. Two choices are the forward and reverse KL divergences, which differ in both their behaviour and the samples required to estimate them. 

Forward KL objectives tend to collapse onto high-density modes and not provide a sufficiently rich learning signal to characterize the full high-dimensional posterior distribution. By contrast, optimizing a reverse KL divergence yields gradients that encourage full posterior coverage and capture the distribution's structure away from the mode.

\subsubsection{Forward KL}
\label{app:fwdKL}
The forward KL divergence is given by
\begin{equation}
\begin{split}
    \mathrm{KL}\bigl(\pi \,\|\, q_\theta\bigr)
    & = \int \pi(x)\,\log\frac{\pi(\boldsymbol{x})}{q_\theta(\boldsymbol{x})}\,\mathrm{d}\boldsymbol{x} \\
    & = \mathbb{E}_{\boldsymbol{x}\sim\pi(\boldsymbol{x})}\left[\log\pi(\boldsymbol{x}) -\log q_\theta(\boldsymbol{x}) \right],
\end{split}
\label{eq:fwdKL_def}
\end{equation}
which requires samples drawn from the target distribution, $\boldsymbol{x}_i \sim \pi (\boldsymbol{x})$. Since $\log \pi(\boldsymbol{x})$ is independent of the flow model parameters, it drops out, yielding the training objective in Eq.~\eqref{eq:fwdKL} via Monte Carlo estimation.

\subsubsection{Reverse KL}
\label{app:revKL}
The reverse KL divergence is
\begin{equation}
\begin{split}
    \mathrm{KL}\bigl(q_\theta \,\|\, \pi\bigr)
    & = \int q_\theta(\boldsymbol{x})\,\log\frac{q_\theta(\boldsymbol{x})}{\pi(\boldsymbol{x})}\,\mathrm{d}\boldsymbol{x} \\
    & = \mathbb{E}_{\boldsymbol{x}\sim q_\theta(\boldsymbol{x})}\left[\log\pi(\boldsymbol{x}) -\log q_\theta(\boldsymbol{x}) \right],
\end{split}
\label{eq:revKL_def}
\end{equation}
which can be estimated using samples generated directly from the model, $\boldsymbol{x}_i \sim q_\theta(\boldsymbol{x})$, leading to Eq.~\eqref{eq:revKL}. This provides direct feedback from the generated samples during training but requires a differentiable simulator to evaluate the log-likelihood and backpropagate the gradients, making the training procedure computationally more demanding.

\begin{figure*}
\vskip 0.02in
\begin{center}
\centerline{\includegraphics[width=1.0\linewidth]{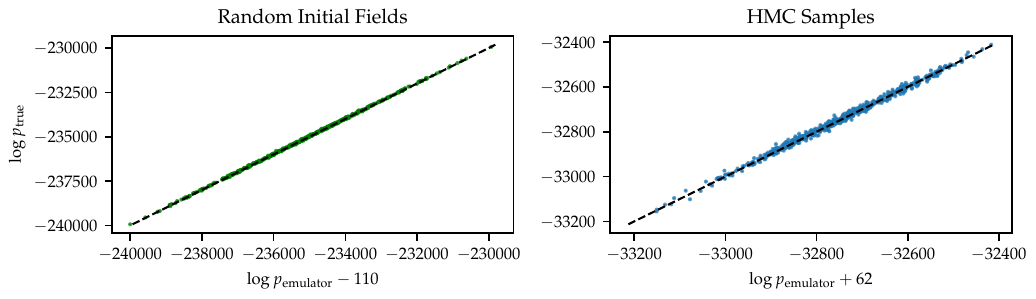}}
\caption{Comparison between un-normalized log posterior values evaluated using Eq.~\eqref{eq:post}, where the forward model is taken to be either the real simulator or the field-level emulator. The left panel shows random test-set samples not used for training, resulting in low-probability values for the given $\boldsymbol{y}_o$. The right panel shows samples drawn from the HMC posterior. Although the emulator is trained only on random input–output pairs, it generalizes to the HMC samples and yields accurate posterior evaluations in the high-probability region. The dashed line indicates perfect agreement; the slightly increased scatter for HMC samples reflects residual emulator errors. Constant offsets are applied for visualization and do not affect relative log-posterior differences between samples.}
\label{fig:surrogate_vs_true}
\end{center}
\vskip -0.2in
\end{figure*}

\subsection{Field-level Emulator for Reverse KL}
\label{app:emulator}
The reverse KL loss in Eq.~\eqref{eq:revKL} requires computing gradients of the forward model with respect to the input fields. This poses a challenge for many scientific simulators, which are often non-differentiable or prohibitively costly to differentiate through. Although the simulator used in this work is differentiable, we include the field-level emulator to replicate the more general case in which gradients of the physical simulator are unavailable. A field-level emulator also has the advantage that it can be evaluated on GPUs, substantially reducing simulation time; in our case, by a factor of $\sim6$ in the forward direction.

\subsubsection{Architecture}
\label{app:FLE_archi}
The field-level emulator network is a lightweight encoder–decoder convolutional V-Net composed of residual 3D convolutional blocks, based on \texttt{map2map}\footnote{\href{https://github.com/eelregit/map2map}{github.com/eelregit/map2map}}. It consists of $2$ down- and upsampling blocks. Each residual block contains two $3^3$ convolutions and a skip connection. Downsampling is implemented with strided $2^3$ convolutions, and upsampling with 3D transposed convolutions of the same stride. Because the convolutions do not use any padding, skip connections from the encoder are symmetrically cropped before concatenation. The input is periodically padded with $20$ voxels on each side to reflect the periodic nature of the cosmological simulation and to yield output fields of size $32^3$. The final output is a single 3D field representing the surrogate LPT prediction. In total, this model has $\sim 1.7$M parameters. 

\subsubsection{Training objective and validation}
The surrogate is trained using a mean-squared error (MSE) loss between predicted and true final density fields,
\begin{equation}
    \mathcal{L}_{\mathrm{MSE}} = \big\lVert \hat{\boldsymbol{y}}_{\mathrm{LPT}} - \boldsymbol{y}_{\mathrm{LPT}} \big\rVert_2^2,
\end{equation}
where $\boldsymbol{y}_{\mathrm{LPT}}$ here is the simulation output without noise and 
$\hat{\boldsymbol{y}}_{\mathrm{LPT}}$ is the prediction by the field-level emulator. While more advanced field-level objectives (e.g.\ particle-displacements instead of gridded density field) could be employed, we find that MSE training alone produces sufficiently accurate predictions for our demonstration.

More specifically, we train on $4096$ random input–output pairs from the simulator using a batch size of $32$ and a learning rate of $10^{-4}$ for $160$ epochs, after which evaluation on a validation set of $1024$ samples shows no further performance improvement.

We validate the field-level emulator by applying it to both randomly sampled initial conditions and to HMC samples. Figure~\ref{fig:surrogate_vs_true} shows that in both cases the predicted final fields, when evaluated using Eq.~\eqref{eq:post}, yield accurate relative probabilities. Notably, the emulator performs comparably on the HMC samples, despite not having been trained on them, indicating good generalization beyond its training data.

\section{Stochastic Interpolant}
\label{app:SI}
Stochastic interpolants \cite{Albergo2025} provide a flexible framework for
constructing diffusion processes that interpolate between two arbitrary
distributions. In the standard diffusion model setting \cite{hoDenoisingDiffusionProbabilistic2020}, one typically defines a path
connecting Gaussian noise $\boldsymbol{x}_0 \sim \mathcal{N}(\boldsymbol{0}, \mathbb{I})$ to a
target variable $\boldsymbol{x}_1$. In our application, however, both endpoints of the
interpolant have clear physical meaning: the observed density field and the
initial conditions. We therefore replace the usual Gaussian endpoint with the
observed field and consider the interpolation
\begin{equation}
    \boldsymbol{x}_0 \equiv  \boldsymbol{y} \; \mathrm{(final \; field)} \;\; \longmapsto \;\; \boldsymbol{x}_1 \equiv \boldsymbol{x} \; \mathrm{(initial \; field)}.
\end{equation}
 
\subsection{Architecture of SI}
We introduce the interpolant
\begin{equation}
    \boldsymbol{I}_s(\boldsymbol{x}_0,\boldsymbol{x}_1,s) = \alpha_s \boldsymbol{x}_0 + \beta_s \boldsymbol{x}_1+\sigma_s W_s(\boldsymbol{z}),
\end{equation}
with $W_s=\sqrt{s}\boldsymbol{z}$ and $\boldsymbol{z}\curvearrowleft \mathcal{N}(\boldsymbol{0},\boldsymbol{I})$. Note that the latent variable $\boldsymbol{z}$ serves different roles in the two methods: in SI it enters as a stochastic drift term, whereas in GLOW it defines the base distribution. We further define $\alpha_s=\sigma_s=1-s$, $\beta_s=s^2$ and with boundary conditions $\alpha_0=\beta_1=1$, and $\alpha_1=\beta_0=\sigma_1=0$. This makes $\boldsymbol{x}_s \equiv \boldsymbol{I}_s(\boldsymbol{x}_0,\boldsymbol{x}_1,s)=(1-s)\boldsymbol{x}_0+s^2\boldsymbol{x}_1+(1-s)W_s(\boldsymbol{z})$. We have that
\begin{equation}
    \boldsymbol{x}_s \curvearrowleft p(\boldsymbol{x}_s|\boldsymbol{x}_0) \quad s\in[0,1].
\end{equation}
The conditional distribution is the solution to SDE to be used as the generative model:
\begin{equation}
    \mathrm{d}\boldsymbol{x}_s = b_s(\boldsymbol{x}_s,\boldsymbol{x}_0)\mathrm{d}s+\sigma_s \mathrm{d}W_s(\boldsymbol{z}) \quad x_{s=0}=\boldsymbol{x}_0.
\end{equation}
The drift $b_s$ is parameterized by a neural network $\hat{b}_s(\boldsymbol{x}_s, \boldsymbol{x}_0, s)$ implemented as a styled V-net based on \texttt{map2map} \citep[as in, e.g.,][]{karrasAnalyzingImprovingImage2020,Jamieson2024}. The architecture is similar to the field-level emulator described in Appendix~\ref{app:FLE_archi}, but extended to a styled setting by conditioning on the interpolant time $s$. This conditioning is introduced via a learned time embedding of dimension $256$, processed by a small multilayer perceptron and injected into the network at each level. Compared to the emulator architecture, this network uses two residual blocks per level, while still using two downsampling and upsampling stages.

\begingroup
\renewcommand{\thefigure}{D1}
\begin{figure*}
\vskip 0.02in
\begin{center}
\centerline{\includegraphics[width=0.99\textwidth]{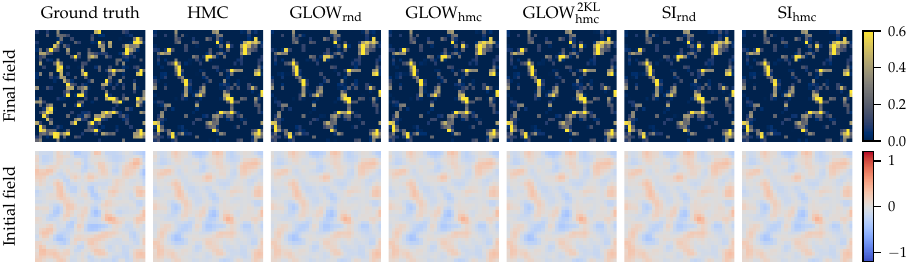}}
\caption{$2$D slices of the $3$D density fields showing the posterior mean reconstructions obtained with the different sampling methods. Mean fields are estimated from samples drawn using HMC, the GLOW model, and the stochastic interpolant (SI), including their respective variants. Comparison with the ground-truth final and initial fields qualitatively demonstrates accurate recovery of the initial conditions. Gaussian smoothing is applied to the initial fields to suppress voxel-scale noise, thereby more clearly revealing correlations on larger scales. Note that perfect recovery of the ground truth is not expected for any method, as the inference is performed on noisy observations obtained by adding a Gaussian noise realization to the ground truth final field (see Fig.~\ref{fig:pk_noise}).}
\label{fig:comp_all}
\end{center}
\vskip -0.2in
\end{figure*}
\endgroup

We want to minimize
\begin{equation}
    \mathcal{L}_{\mathrm{SI}}[\hat{b}_s] = \int_0^1 \mathrm{d}s \mathbb{E}[|\hat{b}_s(\boldsymbol{I}_s,\boldsymbol{x}_0,s)-\boldsymbol{R}_s|^2],
\end{equation}
where
\begin{equation}
\begin{split}
    \boldsymbol{R}_s & = \dot{\boldsymbol{I}_s} = \dot{\alpha_s} \boldsymbol{x}_0 + \dot{\beta_s}\boldsymbol{x}_1 + \dot{\sigma_s}W_s = \\
    & = -\boldsymbol{x}_0 + 2s\boldsymbol{x}_1 - W_s.
\end{split}
\end{equation}
In practice, the integral over $s$ is approximated with Monte Carlo by draws of $s \sim U([0,1])$, which gives Eq.~\eqref{eq:si_loss}.

The Fokker--Planck equation guarantees that the distribution of $\boldsymbol{I}_s$ evolves correctly along this path, ensuring that sampling trajectories initialized at $\boldsymbol{x}_0$ yield samples from the correct distribution of the initial conditions conditioned on the data. This provides a principled and probabilistically valid mechanism for sampling $\boldsymbol{x}$ given $\boldsymbol{y}$. However, this guarantee is exact only under perfect learning of the drift, which in practice corresponds to the idealized limit of infinite training data, sufficient model capacity, and full convergence during training.

During sampling, we use $100$ update steps to interpolate from the final field to the initial field. As in GLOW, periodic padding is applied in the SI within each update step, specifically to the inputs of the drift network.

\subsection{Training}
\label{app:si_training}
The SI models contain $8.8$ million parameters. During both training and sampling, fields are padded by $10$ voxels in each dimension at every intermediate step. Training is performed with a batch size of $32$ on a single NVIDIA A100 GPU with $40\,\mathrm{GiB}$ of memory. We use a learning rate of $10^{-4}$ and run the training for $100$ epochs. We observe no significant performance differences between batch sizes of $16$ and $32$, indicating limited improvement to increased batch size. Scaling to larger models and multi-GPU training will be explored in future work. 

\section{Ground truth vs Posterior means}
\label{app:additional}
Figure~\ref{fig:comp_all} shows posterior mean predictions for both the initial and the corresponding forward-simulated final fields, compared to the ground truth, for samples drawn from HMC, the different GLOW variants, and the SI model. To facilitate comparison with the ground truth, we apply Gaussian smoothing with $\sigma_{\mathrm{smooth}}=1$ voxel, which suppresses voxel-scale scatter while preserving larger-scale structure.

\section{Different ground truths}
\label{app:diff_ground_truths}
To ensure that the results presented in Section~\ref{sec:results} are not specific to a single realization of the ground truth initial conditions and corresponding mock data, we generate two additional mock final density fields. One advantage of the generative models, when trained in the \textit{rnd} setting, is their amortized nature, which enables efficient sampling for new ground truth realizations without retraining. However, for comparison with the target posterior obtained via HMC, new Markov chains must be generated for each case. Subsequent retraining of models on \textit{hmc} samples for these additional realizations requires additional compute, and we therefore do not pursue this here. Instead, we take the agreement with the base \textit{rnd} case as support for the robustness of the results across realizations, and expect the observed improvements from enhanced training data and loss functions to generalize.

For each new ground truth, we run a single HMC chain using \texttt{BORG}. Each chain produces $0.66\times10^6$ samples, and we assess convergence and estimate the number of independent samples following the procedure described in Appendix~\ref{app:borg}. We find comparable autocorrelation lengths across chains and adopt a thinning lag of $50$, resulting in $13\,160$ independent samples per ground truth, which define the corresponding HMC posteriors. To limit computational overhead, we restrict the analysis to this number of samples, which we find sufficient to assess robustness across different ground truths. For consistency, we generate an equal number of samples from the trained GLOW and SI models (from the \textit{rnd} setting). 

\begin{figure}
    \centering
    \includegraphics[width=1.\linewidth]{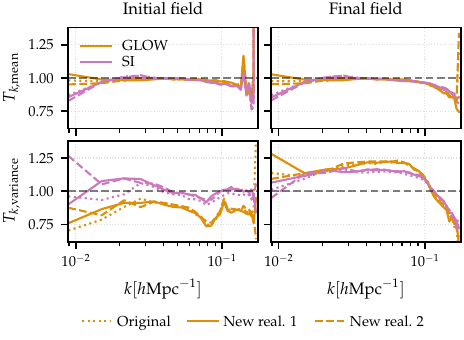}
    \vspace{-2em}
    \caption{Power-spectrum ratios relative to HMC; as Fig.~\ref{fig:powspec}. \textbf{Top:} the ratio for the posterior mean, probing the recovery of spatial structure across scales. \textbf{Bottom:} the ratio for the variance, quantifying spatial correlations in posterior uncertainty. Dotted, solid, and dashed lines correspond to the original and two additional ground truth realizations, respectively.}
    \label{fig:powratio_new_real}
\end{figure}

\begin{figure}
    \centering
    \includegraphics[width=1.\linewidth]{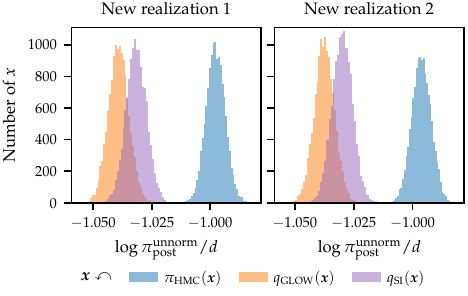}
    \vspace{-2em}
    \caption{Histograms of the unnormalized log posterior evaluated under the target posterior (as in Fig.~\ref{fig:priorlhpost}) for two additional ground truth realizations, comparing HMC, GLOW, and SI. We see consistent behavior across realizations.}
    \label{fig:post_new_real}
\end{figure}

For both additional ground truth realizations, we evaluate the samples from HMC, GLOW, and SI using all diagnostics presented in Section~\ref{sec:results}, and find consistent qualitative behavior across all cases. Representative diagnostics are shown in Fig.~\ref{fig:powratio_new_real} and Fig.~\ref{fig:post_new_real}.

\bsp	
\label{lastpage}
\end{document}